\documentclass[prb,aps,twocolumn,superscriptaddress]{revtex4-1}
\usepackage{graphicx,color}
\usepackage{amsthm}
\usepackage{amsfonts}
\usepackage{algorithmic}
\usepackage{enumerate}
\usepackage{latexsym}
\usepackage{amsmath}
\usepackage{amssymb}
\usepackage{bm}
\usepackage[pdftex,plainpages=false,colorlinks=true,linkcolor=blue, citecolor=blue, urlcolor=blue]{hyperref}

\emergencystretch=\maxdimen
\begin{document}
\title{Phase diagram of the Hubbard model on a square lattice: \\A cluster slave-spin Study}

\author{Ming-Huan Zeng}
\affiliation{Department of Physics, Beijing Normal University, Beijing 100875, China}
\author{Tianxing Ma}
\email{txma@bnu.edu.cn}
\affiliation{Department of Physics, Beijing Normal University, Beijing 100875, China}
\author{Y.-J. Wang}
\email{yjwang@bnu.edu.cn}
\affiliation{Department of Physics, Beijing Normal University, Beijing 100875, China}

\date{\today}

\begin{abstract}
The cluster slave-spin method is employed to investigate systematically the ground state properties of the Hubbard model
on a square lattice with doping $\delta$ and coupling strength $U$ being its parameters. At half-filling, a relation
between the staggered magnetization $M$ and the antiferromagnetic (AFM) gap $\Delta_{\text{AFM}}$ is established
in the small $U$ limit to compare with that from the Hartree-Fock theory,
and a first-order metal-insulator Mott transition in the
paramagnetic state is substantiated, which is characterized by discontinuities and hystereses at $U_{\text{Mott}}=10t$.
The interaction $U_{c}$ for the crossover in the AFM state, separating the weak- and strong-coupling regimes, is found
to remain almost unchanged with large dopings, and smaller than $U_{\text{Mott}}$ at half-filling because of long range
AFM correlations. Finally, an overall phase diagram in the $U$-$\delta$  plane is presented, which is composed
of four regimes: the AFM insulator at half-filling, the AFM metal with the compressibility
$\kappa>0$ or $\kappa<0$, and the paramagnetic metal, as well as three phase transitions: (i) From the AFM
metal to the paramagnetic metal, (ii) between the AFM metal phases with positive and negative $\kappa$,
and (iii) separating the AFM insulating phase at $\delta=0$ from the AFM metal phase
for $\delta>0$.
\end{abstract}


\maketitle

\section{INTRODUCTION}

Because of its intimate relation to the high-$T_{c}$ unconventional superconductivity based on cuprate oxides, the one-band
Hubbard model~\cite{Hubbard1963, *Hubbard1964} on a square lattice  has been extensively studied through many theoretical
approaches --- Green's function methods~\cite{Hubbard1963,Hubbard1964,PhysRevLett.23.1448,Cryot1972,PhysRevB.96.081117},
slave-variable representations~\cite{PhysRevLett.57.1362, PhysRevB.76.155102, PhysRevB.81.035106, PhysRevB.86.085104,
PhysRevB.96.115114,PhysRevB.49.2368,*Feng_2004, *Feng_2015}, variational methods
for the wave-functions and spin configurations~\cite{PhysRev.137.A1726,TohruOgawa,Takano1975,CHAO1974525,RevModPhys.56.99},
renormalization-group methods~\cite{PhysRevB.25.6748,RevModPhys.66.129,ZPB.103.339,JPSJ.73.1251}, and numerical methods such as
the Quantum Monte Carlo (QMC)
simulations~\cite{PhysRevB.31.4403,PhysRevB.94.195126,PhysRevB.94.085140,PhysRevB.80.075116,PhysRevLett.51.1900,
PhysRevB.38.11809,PhysRevLett.124.117602,PhysRevB.78.165101,PhysRevB.91.241116}, cluster dynamical
mean-field theory  (CDMFT)~\cite{PhysRevLett.101.186403,PhysRevB.95.235109,PhysRevB.96.241109}, density matrix embedding
theory (DMET)~\cite{Zheng2016,ZhengBX2017}, etc. Up till now,  some consensuses about this model  have been reached, e.g., the
first-order metal-insulator Mott transition in the half-filled paramagnetic (PM)
state~\cite{RevModPhys.68.13,RevModPhys.77.1027,PhysRevLett.87.167010,PhysRevLett.90.099702,PhysRevLett.101.186403,
JPSJ.80.084705,JPSJ.73.1251} and  an infinitesimal critical coupling strength for the antiferromagnetic (AFM) phase at
half-filling because of the nested Fermi surface in the square
lattice~\cite{PhysRevLett.23.1448,Cryot1972,PhysRevB.31.4403,PhysRevLett.57.1362,PhysRevB.91.125109,
PhysRevLett.124.017003,PhysRevB.40.506,PhysRevB.99.165121,PhysRevLett.62.591,PhysRevB.94.195126,
PhysRevB.94.085140,PhysRevB.80.075116,JPSJ61,Zheng2016}. Even if the Mott transition in the half-filled PM system transforms
into a crossover in the AFM state~\cite{PhysRevB.94.085140,PhysRevLett.124.117602,PhysRevB.94.195126},  the relationship
between them has not yet been understood thoroughly.
There are two types of phase separation: the one with hole-rich and hole-poor
regions~\cite{PhysRevB.78.165101,PhysRevB.91.241116,PhysRevB.94.195126} and that with incommensurate
AFM~\cite{PhysRevLett.64.1445} at small and intermediate $U$. While the earlier
works~\cite{PhysRev.142.350,PTP.99.489,EPJB.27.473} present the phase separation as a mixture of AFM and FM states in the
large $U$ limit, we find that the phase separation happens easier when doping the AFM Slatter insulator.
It is worthy of  noticing that the theoretical and numerical results can not be compared with experiments on the cuprate
oxides directly because some crucial simplifications have been made in the Hubbard model by neglecting the long-range
Coulomb interaction, the hopping matrix elements further than the nearest neighbors,  and the inter-layer correlations,
which have made the phase diagram of the two-dimensional (2D) Hubbard model~\cite{PhysRevB.96.081117,PhysRevB.94.195126}
qualitatively different from that of high-$T_{c}$ cuprate superconductors~\cite{RevModPhys.78.17}.  However, recent
experimental improvement on the ultracold atoms to lower the local temperature of the optical lattice beyond the exchange
energy $J$ make it possible to observe the spatial charge and spin correlations even beyond the nearest
neighbors~\cite{Daniel2016,Martin2016,Cheuk2016}. Thus, the optical lattice platform could not only help us unveil the
phase diagram in the intriguing regime where the kinetic energy and interaction potential become comparable, but also
examine the validity of various theoretical methods based on different approximations. Meanwhile, more reliable results
of the model ought to be needed to provide guidelines for experimental researches.

Although the QMC method have succeeded in coping with the half-filled bipartite lattice, the sign problem for the doped
fermion systems and the finite size effect make its predictions for non-half-filled systems at least questionable. In
these cases, some nonperturbative semi-analytic methods based on slave-variable representations have  been introduced
to throw some light on the regime with modest coupling strengths, where the conventional perturbative techniques fail
to give rise to correct solutions. Among all the slave-particle approaches, slave-boson method has been applied to the
single band~\cite{PhysRevLett.57.1362} and multiorbital systems successfully~\cite{PhysRevB.76.155102}. Earlier, Kotliar
and Rukenstein have proven that~\cite{PhysRevLett.57.1362} the saddle point approximation within the slave-boson approach
could reproduce the Gutzwiller approximation's results of the half-filled  PM state, and high-energy excitations can be
taken into account by the fluctuations around the saddle point. Nevertheless, as  the dimension of degrees of freedom
increases,  the exponentially growing number of bosonic slave variables makes the slave boson approach particularly
inapplicable to multiorbital systems and cluster mean-field theories.  For the large $U$ Hubbard model, a fermion-spin
transformation~\cite{PhysRevB.49.2368} has been developed to implement the charge-spin separation in its equivalent,
$t$-$J$ model, where the charge degree of freedom is represented by a spinless fermion while the spin degree of freedom
by a hard-core boson. Within this scheme and its development, both the normal state properties and superconductivity
mechanism of doped cuprates has been investigated extensively~\cite{PhysRevB.68.184501}.

Recently, the slave-spin approach~\cite{PhysRevB.70.035114,PhysRevB.76.195101,PhysRevB.66.165111,PhysRevB.72.205124}
has been devised to study the half-filled multiorbital Hubbard model, where the slave spin is introduced to represent
the charge sector of an electron. The difficulty mentioned above has been overcome because only one slave variable is
needed for each degree of freedom.  After that, the slave-spin method was improved by Hassan and
de' Medici~\cite{PhysRevB.81.035106} to deal with the non-half-filled systems, known as the $Z_{2}$ slave-spin theory.
However, it has been proven that the $Z_{2}$ slave-spin theory fails to reproduce the noninteracting dispersion of the
multi-orbital Hamiltonian because of the non-zero orbit-dependent Lagrange multipliers. The $U(1)$ slave-spin
theory~\cite{PhysRevB.86.085104} can not only produce the same results as the $Z_{2}$ theory's when the on-site
interaction is on, but also capture the correct dispersion in the noninteracting limit by the extra orbit-dependent
effective chemical potential, which is identical to the Lagrange multiplier at $U=0$
[see Appendix.~\ref{Single-Site-Approximation} for details].  A recently generalized cluster slave-spin
method~\cite{PhysRevB.96.115114} has succeeded in describing the crossover of the AFM  gap $\Delta_{\text{AFM}}$
and obtaining the correct  quasi-particle residue that is consistent with the extended Gutzwiller
factor~\cite{TohruOgawa}. However, there remain some insufficiencies:
(i) The solutions at $U<5t$ are absent.
(ii) They have not discussed the relationship between the crossover of the AFM gap and the Mott transition in the
half-filled PM state.
(iii) The properties of the charge component represented by the slave-spin variables have not been studied on the
same footing as the spin degree of freedom.
(iv) The phase diagram in the $U$-$\delta$ plane has not been given, which can provide us a holistic understanding
of the ground state properties.
(v) No attention has been paid to the half-filled system.
By completing the missing parts mentioned above, we offer an overall phase diagram in the $U$-$\delta$ plane that
consists of four regimes: AFM insulator, AFM metallic phases with positive and negative compressibility, and the
PM metal. The first-order Mott transition occurs in the half-filled PM state, characterized by discontinuities and
hystereses at $U=10t$, which transforms into an extended crossover in the AFM state because of nonlocal AFM
correlations. Besides, the phase separation, manifested by the negative compressibility, has been observed in
the regions with intermediate dopings, indicating that the uniform AFM state is not the true ground state of the model.

The rest of this paper is organized as follows.
In Sec.~\ref{FORMALISM}, we outline the cluster slave-spin mean-field theory of Ref.~\onlinecite{PhysRevB.96.115114}.
In Sec.~\ref{FINITEDOPE},  the results of hole-doped systems obtained by the 2/4-site cluster approximations are
discussed comprehensively.
In Sec.~\ref{HALF-FILL}, for the half-filled system, we establish an analytical relation between $M$ and
$\Delta_{\text{AFM}}$ in the small $U$ limit, and observe the first-order Mott transition in the PM state.
In Sec.~\ref{PHASES}, $M$, $\Delta_{\text{AFM}}$, and the compressibility $\kappa$ as functions of $U$ and $\delta$
are studied thoroughly, and an overall phase diagram in the $U$-$\delta$ plane is presented.

\section{Formalism}\label{FORMALISM}

We start from the standard single-band fermionic Hubbard model~\cite{Hubbard1963,Hubbard1964} defined by
\begin{equation}\label{Hubbard model}
H = -t \sum_{\langle i,j \rangle \sigma} (c_{i\sigma}^\dagger c_{j\sigma} + \mbox{\sc h.c.}) + U \sum_{i} n_{i\uparrow}
n_{i\downarrow} - \mu \sum_{i\sigma} n_{i\sigma} \;,
\end{equation}
where $t$, $U$, $\mu$ are the nearest hopping constant, the on-site Coulomb repulsion energy and the chemical potential,
respectively. $\langle i,j \rangle $  represents that the sum is over the nearest neighbors, and
$c_{i\sigma}(c_{i\sigma}^\dagger )$ annihilates (creates) an electron at site $i$ with spin
$\sigma= \uparrow,\, \downarrow$, and the number operator
$n_{i\sigma} = c_{i\sigma}^\dagger c_{i\sigma}$.

\begin{figure}[htb!]
\centering{
\includegraphics[
scale=0.16]{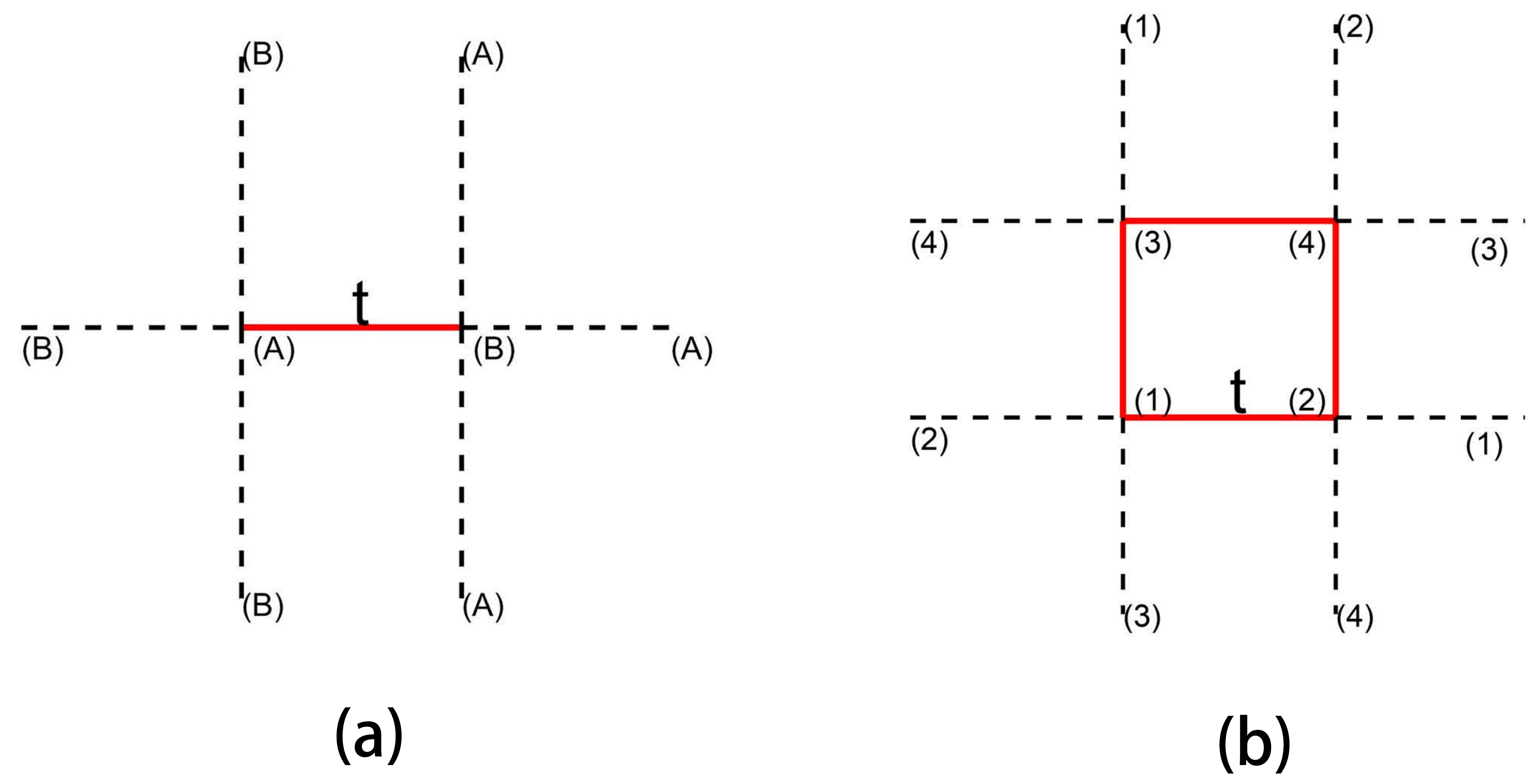}
}
\caption{Schematic illustration of the (a) two- and (b) four-site cluster}
\label{twofour}
\end{figure}

In the slave-spin method, the electron operator $c_\alpha$ is decomposed into the product of a fermionic spinon and
a slave spin with $S=\tfrac{1}{2}$  which represents the physical spin and charge degrees of freedom, respectively,
\begin{equation}
c^{\dagger}_{\alpha}=S^{\dagger}_{\alpha}f^{\dagger}_{\alpha}\;.
\end{equation}
With the constraint $a^{\dagger}_{\alpha}a_{\alpha}+b^{\dagger}_{\alpha}b_{\alpha}=1$, the slave-spin operator is
rewritten in the  Schewinger boson representation
\begin{equation}\label{SSPIN VARIABLE}
S^{\dagger}_{\alpha}=a^{\dagger}_{\alpha}b_{\alpha} \;, \;\;\;\;
S^{z}_{\alpha}=\frac{1}{2}\left(a^{\dagger}_{\alpha}a_{\alpha}-b^{\dagger}_{\alpha}b_{\alpha}\right)\;,
\end{equation}
which has been utilized to describe the metal-insulator Mott transition in the multiorbital Hubbard
model~\cite{PhysRevB.86.085104}. In the slave-spin formalism, the original Hillbert space of a single degree
of freedom with the basis $\{|n_{\alpha}\rangle\}=\{|0\rangle,|1\rangle\}$ is enlarged to
$\{|n^{f}_{\alpha},S^{z}_{\alpha}\rangle\}=\{|0,-\tfrac{1}{2}\rangle,|1,\tfrac{1}{2}\rangle,
|0, \tfrac{1}{2}\rangle,|1,-\tfrac{1}{2}\rangle\}$. Thus, the following constraint is enforced to project out the
unphysical states $|0,\tfrac{1}{2}\rangle,|1,-\tfrac{1}{2}\rangle$:
\begin{equation}\label{constraint}
S^{z}_{\alpha}=f^{\dagger}_{\alpha}f_{\alpha}-\tfrac{1}{2}\;.
\end{equation}
However, according to Kotliar and Ruckenstein~\cite{PhysRevLett.57.1362}, the slave-spin operators defined
in Eq.~\eqref{SSPIN VARIABLE} need to be dressed to  ensure the correct behavior in the noninteracting limit
\begin{eqnarray}
\tilde{S}^{\dagger}_{\alpha}&=&P^{\dagger}_{\alpha}a^{\dagger}_{\alpha}b_{\alpha}P_{\alpha}\;,\nonumber\\
P^{\pm}_{\alpha}&=&\frac{1}{\sqrt{1/2 \pm S^{z}_{\alpha}}}\;.
\end{eqnarray}
Then, in the slave-spin method, the Hamiltonian (\ref{Hubbard model}) can be rewritten as
\begin{eqnarray}\label{Hubbard-SSpin-F}
 H&=&-t \sum_{\langle i,j\rangle \sigma} (\tilde{S}_{i\sigma}^+ f_{i\sigma}^\dagger\tilde{S}_{j\sigma}^- f_{j\sigma}
 + \textsc{h.c.} )  -\mu \sum_{i\sigma} f_{i\sigma}^\dagger f_{i\sigma}\nonumber\\
&& - \sum_{i\sigma} \lambda_{i\sigma} (f_{i\sigma}^\dagger f_{i\sigma} - S_{i\sigma}^z - \tfrac{1}{2} )\nonumber\\
&&+\frac{U}{2}  \sum_{i\sigma} (S^{z}_{i\sigma} +\tfrac{1}{2})(S^{z}_{i-\sigma} +\tfrac{1}{2})\;,
\end{eqnarray}
where the constraint (\ref{constraint}) is taken care of by introducing a Lagrange multiplier $\lambda_{i\sigma}$.
Under the two-site cluster approximation sketched in Fig.~\ref{twofour}(a), Eq.~\eqref{Hubbard-SSpin-F} can be decomposed
into a fermionic spinon Hamiltonian
and a cluster slave-spin Hamiltonian as
\begin{eqnarray}
H^f_{\text{MF}}& =& \sum_{\langle i,j\rangle \sigma} \big[ -tZ-\delta_{ij}(\mu+\lambda_{I\sigma} -\tilde{\mu}_{I\sigma}) \big]
f_{i\sigma}^\dagger f_{j\sigma} + \textsc{h.c.}\;, \nonumber\\
\label{FERMI-HAML}\\
H_{\text{2-site}}^{S} &=& \sum_{\sigma}\big[\lambda_{\sigma}^{+}(S_{A\sigma}^z + S_{B\sigma}^z)+\lambda_{\sigma}^{-}
(S_{A\sigma}^z - S_{B\sigma}^z)\big] \nonumber\\
 &&+\sum_{\sigma} \Big\{ \epsilon_{\sigma}^{x} ( \tilde{z}_{A\sigma}^\dagger \tilde{z}_{B\sigma}
+\tilde{z}_{B\sigma}^\dagger \tilde{z}_{A\sigma} )\nonumber\\
&&+ (\epsilon_{\sigma}^{x} + 2 \epsilon_{\sigma}^{y} ) \big[
\tilde{z}_{A\sigma}^\dagger \langle \tilde{z}_{B\sigma} \rangle
+ \tilde{z}_{B\sigma}^\dagger \langle \tilde{z}_{A\sigma} \rangle + \textsc{h.c.} \big] \Big\} \nonumber \\
&&+ U  (S_{A\uparrow}^z + \tfrac{1}{2}) (S_{A\downarrow}^z + \tfrac{1}{2})
+ U (S_{B\uparrow}^z + \tfrac{1}{2}) (S_{B\downarrow}^z + \tfrac{1}{2}) \;.
\label{SSPIN-HAML-2SITE}\nonumber\\
\end{eqnarray}
where $I=A/B$, and
\begin{eqnarray}
\hspace{-2em} Z&=&\langle\tilde{z}_{A\sigma}^\dagger\rangle\langle\tilde{z}_{B\sigma}\rangle,\;\hspace{3.2em}
\lambda_{\sigma}^{\pm} =\frac{\lambda_{A\sigma}\pm \lambda_{B\sigma}}{2},\nonumber\\
\hspace{-2em} \tilde{\mu}_{I\sigma}&=& \frac{4 Z \langle S_{I\sigma}^z \rangle (\epsilon_\sigma^x
+ \epsilon_\sigma^y) }{( \tfrac{1}{2} )^2 - \langle S_{I\sigma}^z \rangle^2 } \;,\;\;\;\;
\epsilon_{\sigma}^{x/y} =-t \langle f_{i\sigma}^\dagger f_{i+\hat{x}/\hat{y}\sigma} \rangle \;.
\end{eqnarray}
The spinon part is readily Fourier transformed into $\bm{k}$-space:
\begin{eqnarray}\label{FERMI-HAML-K}
H^f_{\text{MF}}
&=& \sum_{\bm{k},\sigma} ( \xi_{\bm{k}\sigma} f_{\bm{k}\sigma}^\dagger f_{\bm{k}\sigma}
+ \Delta_\sigma f_{\bm{k}\sigma}^\dagger f_{\bm{k} + \bm{Q}\sigma} ) \;,
\end{eqnarray}
where
\begin{eqnarray}
\xi_{\bm{k}\sigma} &=& -4t Z \gamma_{\bm{k}} - \mu_\sigma^{\text{eff}}\;, \hspace{1.8em} \gamma_{\bm{k}}
= \tfrac{1}{2} (\cos k_x + \cos k_y) \;, \nonumber\\
\mu^{\text{eff}} &=& \mu - \tfrac{1}{2} (\tilde{\mu}_{A\sigma} - \lambda_{A\sigma} + \tilde{\mu}_{B\sigma}
- \lambda_{B\sigma}) \;, \nonumber\\
\Delta_\sigma &=& \tfrac{1}{2} (\tilde{\mu}_{A\sigma} - \lambda_{A\sigma} - \tilde{\mu}_{B\sigma}
+ \lambda_{B\sigma}) \;. \label{defDLTAFM}
\end{eqnarray}
Noticing that $\Delta_\sigma=-\Delta_{-\sigma}$ in the AFM state, the PM state can be reached by simply forcing
the AFM gap $\Delta_{\text{AFM}}=|\Delta_{\sigma}|=0$.

Still, there are two key points needed to be cleared.
First, the fermionic spinon Hamiltonian~\eqref{FERMI-HAML} is very similar to that in the Hartree-Fock (HF)
approximation except the hopping constant $t$ is
renormalized by the quasi-particle residue $Z$  as $tZ$. Second, the Hamiltonian~(\ref{SSPIN-HAML-2SITE}) in the
slave-spin sector is a repulsively interacting Bose-Hubbard model for bosons $a_{i\sigma}$ and $b_{i\sigma}$
in the staggered external field $\lambda_{A\sigma}\;\text{and}\;\lambda_{B\sigma}$.  When the composite boson fields
$\tilde{z}_{i\sigma}$ condensate, i.e., $\langle \tilde{z}_{i\sigma}\rangle\neq0$ below the critical coupling strength
$U_{\text{Mott}}$ for the metal-insulator transition in the PM state, the system changes from the insulating phase to
a metallic one~\cite{PhysRevB.86.085104}. Moreover, when $U$ is strong enough to make
$\lambda_{A\sigma}\neq \lambda_{B\sigma}$, i.e., $\lambda_{\sigma}^{-}\neq0$, the system will transit into the AFM state
from the PM state. Thus, the Hamiltonian~ \eqref{FERMI-HAML} and (\ref{SSPIN-HAML-2SITE}) can describe both the
Mott transition at $U=U_{\text{Mott}}$ in the PM phase and the PM-to-AFM transition at $U_{M}$.

It has been proven that the four-site cluster approximation makes a great improvement compared to the two-site one by
including more inter-site fluctuations~\cite{PhysRevB.96.115114}. Thus, except Figs.~\ref{SSPINdope002} and
\ref{dope00-2-4site}, all our results are obtained in the four-site cluster approximation as illustrated in
Fig.~\ref{twofour}(b), whose mean-field fermionic spinon Hamiltonian remains the same as Eq.~\eqref{FERMI-HAML},
and the cluster slave-spin Hamiltonian reads:
 \begin{eqnarray}\label{SSPIN-HAML}
 H_{\text{4-site}}^{S}&=&H^S_{\lambda}+ H^S_{U}+ H^S_{K} \;,
 \end{eqnarray}
 where
 \begin{eqnarray}
 H^S_{\lambda} &=& \sum_{I=1\sigma}^4\lambda_{I\sigma}S^z_{I\sigma} \\
 H^S_{U} &=& \sum_{I}U  (S_{I\sigma}^z + \tfrac{1}{2}) (S_{I \bar{\sigma}}^z + \tfrac{1}{2})\label{SSPIN-POT}\\
 H^S_{K} &=& \sum_{\sigma} \Big\{ \epsilon_{\sigma}^{x}(\tilde{z}_{1\sigma}^\dagger \tilde{z}_{2\sigma}
 +\tilde{z}_{3\sigma}^\dagger \tilde{z}_{4\sigma})
 + \epsilon_{\sigma}^{y}(\tilde{z}_{1\sigma}^\dagger \tilde{z}_{3\sigma}
 +\tilde{z}_{2\sigma}^\dagger \tilde{z}_{4\sigma})\nonumber\\
&& \;\;\;\; + \epsilon_{\sigma}^{x} \big[\tilde{z}_{1\sigma}^\dagger \langle \tilde{z}_{2\sigma} \rangle
+ \tilde{z}_{2\sigma}^\dagger \langle \tilde{z}_{1\sigma} \rangle+
\tilde{z}_{3\sigma}^\dagger \langle \tilde{z}_{4\sigma} \rangle+ \tilde{z}_{4\sigma}^\dagger
\langle \tilde{z}_{3\sigma} \rangle \big]\nonumber\\
&&\;\;\;\; +\epsilon_{\sigma}^{y} \big[\tilde{z}_{1\sigma}^\dagger \langle \tilde{z}_{3\sigma} \rangle
+ \tilde{z}_{3\sigma}^\dagger \langle \tilde{z}_{1\sigma} \rangle +
\tilde{z}_{2\sigma}^\dagger \langle \tilde{z}_{4\sigma} \rangle+ \tilde{z}_{4\sigma}^\dagger
\langle \tilde{z}_{2\sigma} \rangle\big] \nonumber\\
&& \;\;\;\; \;\;\;\; + \textsc{h.c.} \Big\} \;. \label{SSPIN-KIN}
\end{eqnarray}
In this work, we only investigate the ground state properties of the system and will adopt the following density
of states in all computations to avoid the finite size effect,
\begin{equation}\label{density}
D(\gamma) = \frac{1}{N} \sum_{\bm{k}} \delta(\gamma -\gamma_{\bm{k}})
= \frac{2}{\pi^2} K(\sqrt{1-\gamma^2}) \; ,
\end{equation}
 where $K(x)$ is the complete elliptical function of the first kind.

\section{RESULTS AND DISCUSSIONS}

\subsection{SYSTEMS WITH FINITE DOPING}\label{FINITEDOPE}

\begin{figure}[htb!]
\centering{
\includegraphics[
scale=0.58]{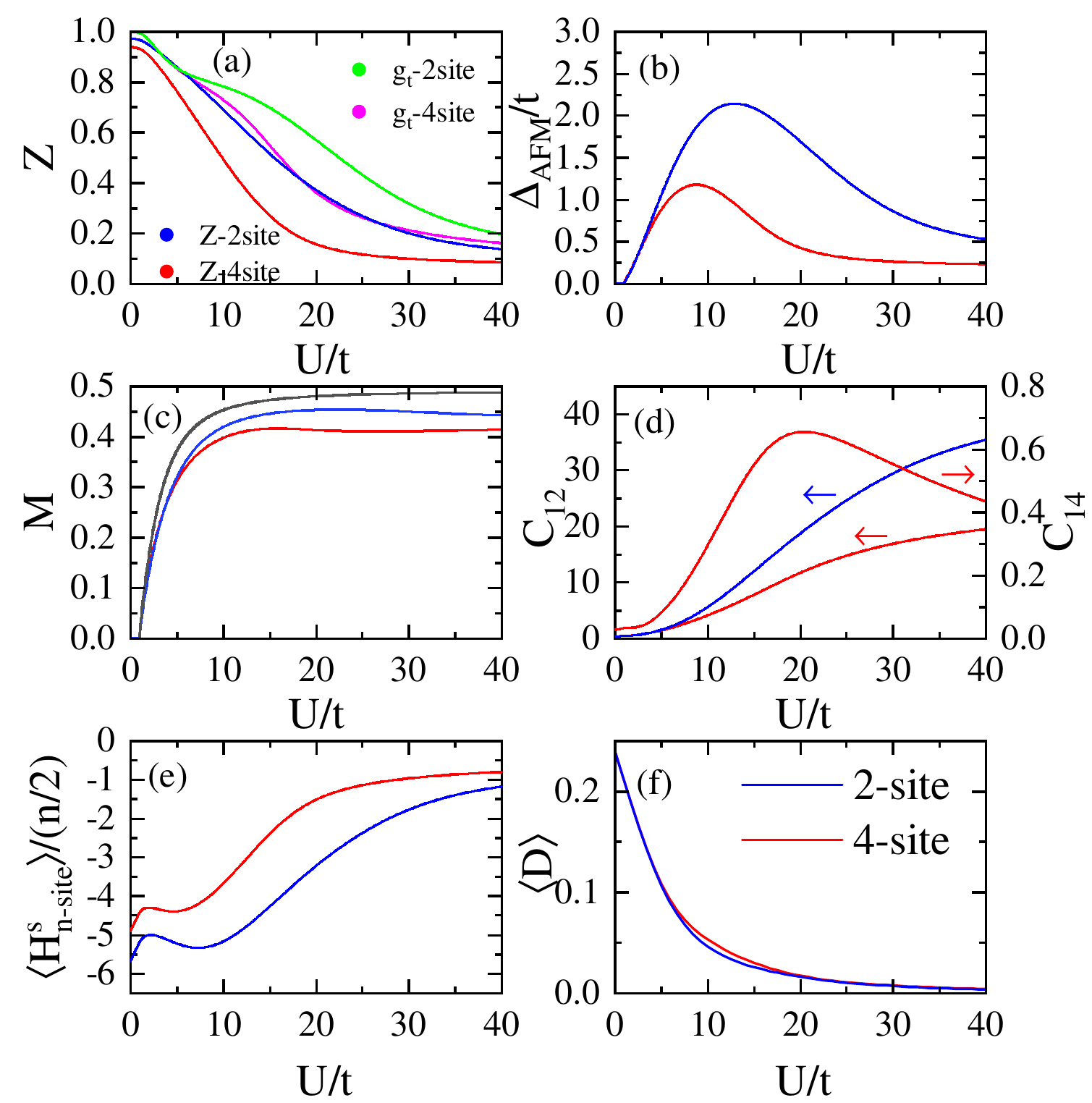}
}
\caption{(a) The quasi-particle weight $Z$ and the generalized Gutzwiller factor $g_t$, (b) the AFM energy gap
$\Delta_{\text{AFM}}/t$, (c) the staggered magnetization $M$,  (d) the holon-doublon correlator between the
nearest neighbors $C_{12}$ and the next nearest neighbors  $C_{14}$, (e) the expectation value of the cluster
slave-spin Hamiltonian $\langle H_{\text{2-site}}^{S}\rangle$ and $\langle H_{\text{4-site}}^{S}\rangle/2$,
and (f) the double occupancy $\langle D\rangle$ as functions of $U$ at $\delta=0.02$  in the AFM state obtained
by the two-site (blue) and the four-site (red) cluster. For comparison, the staggered magnetization within
the HF theory at $\delta=0.02$ is plotted in Fig.~\ref{SSPINdope002}(c) as black line.}
\label{SSPINdope002}
\end{figure}

We analyze the quasi-particle residue $Z$, the generalized Gutzwiller factor
$g_t$~\cite{TohruOgawa,PhysRevB.96.115114,PhysRevB.88.094502}, the AFM gap $\Delta_{\text{AFM}}$, the staggered
magnetization $M$, the holon-doublon correlators between the nearest neighbors $C_{12}$ and the next nearest
neighbors $C_{14}$ [cf. (\ref{ho_do_co}) for definition], the ground state energy of the slave-spin Hamiltonian
$\langle H_{\text{2-site}}^{S}\rangle$, $\langle H_{\text{4-site}}^{S}\rangle/2$, and the double occupancy
$\langle D \rangle$ as functions of $U$ at $\delta=0.02$ obtained from the two- and four-site cluster
approximations in Fig.~\ref{SSPINdope002}.

\begin{figure*}[htb!]
\centering{
\includegraphics[
scale=0.48]{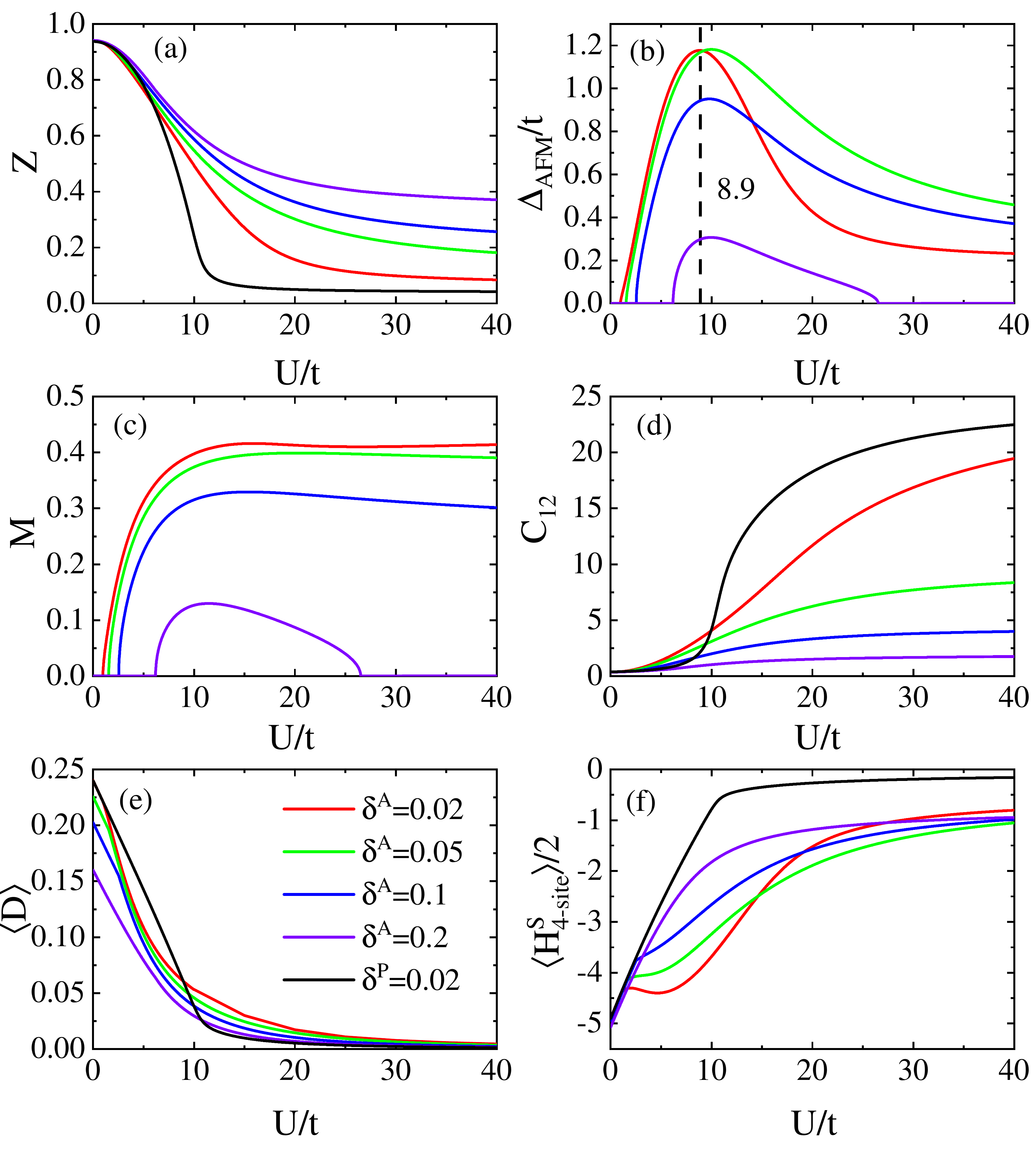}
}
\caption{(a) The quasiparticle weight $Z$, (b) the AFM energy gap $\Delta_{\text{AFM}}/t$, (c) the staggered
magnetization $M$,  (d) the holon-doublon correlator between the nearest neighbors $C_{12}$, (e) the double
occupancy per site, and (f) the expectation value of the cluster slave-spin Hamiltonian as functions of
 $U$  at a set of  doping concentrations $\delta=$ 0.02 (red), 0.05 (green), 0.1 (blue), 0.2 (violet) in the
 AFM state and $\delta=$ 0.02 (black) in the PM state obtained by the four-site cluster.}
\label{dope-vari-4site}
\end{figure*}

The results are as follows:
(i) In Fig.~\ref{SSPINdope002}(a), we find that $Z$'s behave similarly with $g_t$ in both the two- and four-site
cluster approximations, and it is suppressed drastically when extra inter-site fluctuations are taken into account
as the cluster size is enlarged from two to four.
(ii) In Fig.~\ref{SSPINdope002}(b), the crossover $U_c$ separating the weak- and strong-coupling regime in the AFM
state is defined by the peak position of $\Delta_{\text{AFM}}$, which decreases from $12.9t$ (2-site)  to $8.9$ (4-site).
Both $\Delta_{\text{AFM}}$ and $M$ [Fig.~\ref{SSPINdope002}(c)] are restrained appreciably by the inter-site fluctuations.
One observes that $M$ from the cluster slave-spin approximation and the HF theory are consistent when $U$ is small.
(iii) In Fig.~\ref{SSPINdope002}(d), a maximum of $C_{14}$ around $U \sim 20t$ can be identified as one of the features
of the strong correlation, the reason for which has been analyzed in Appendix \ref{Correlator}.
(iv) In  Fig.~\ref{SSPINdope002}(e), $\langle H_{\text{n-site}}^{S}\rangle/(n/2)$ in the AFM states declines suddenly
at $U\sim U_{M} \sim t$, which is caused by the decrease of the interaction potential compared to the PM state in
this region [cf. Fig.~\ref{dope-vari-4site} (f)], and behaves as $-1/U$ at large $U$, which is similar to
$\Delta_{\text{AFM}}/t$.
(v) In  Fig.~\ref{SSPINdope002}(f), as $U$ goes up, the double occupancy decreases monotonically, which probably
tends to zero as $U$ goes to infinity. Finally, we find that the difference between the results from two
approximations is much smaller in the small and large $U$ limit than those at intermediate $U$'s, which
necessitates enlarging the cluster size to study the properties of Hubbard model with moderate coupling strengths.

To understand how the quantities discussed in Fig.~\ref{SSPINdope002} evolve with the increase of $U$ at different doping levels,
their dependence on $U$ at a set of dopings $\delta^P=0.02,\;\delta^A=0.02,\,0.05,\,0.1,\,0.2$ are plotted in
Fig.~\ref{dope-vari-4site}, where A and P denote the AFM or PM  state, respectively.
First, in Fig.~\ref{dope-vari-4site}(a), we find that $Z$ in the PM state is larger than its counterpart in the
AFM state when $U<6t$, or otherwise when $U>6t$, and the quasi-particle weight in the PM state decreases more
dramatically than that in the AFM state around $U\sim U_{\text{Mott}} = 10t$ at $\delta=0.02$. Moreover, the
increase of $Z$ with doping in the AFM state is observed at all $U$'s, signifying the system's tendency towards
a correlated metal.
Second, the $\Delta_{\text{AFM}}$ at all dopings present a crossover behavior at $U_c$ and then reaches a finite
value except for $\delta=0.2$ whose $\Delta_{\text{AFM}}$ vanishes at $U=26t$ in Fig.~\ref{dope-vari-4site}(b).
The reason for this unusual behavior may be that at $\delta=0.2$, the system is an AFM metal, whereas the increased
$U$ makes the system more localized which is unfavorable to the itinerant AFM. It is worth mentioning that
$\Delta_{\text{AFM}}$ shows a nonmonotonic behavior with the increase of doping when $U>U_{\text{Mott}}$.
Third, as shown in Fig.~\ref{dope-vari-4site}(c), $M$ decreases monotonically with doping and like
$\Delta_{\text{AFM}}$ at $\delta=0.2$, the staggered magnetization also vanishes after reaching its maximum
at $U_c$. The same picture has been observed in the recent DMET calculation~\cite{Zheng2016} and some earlier
work by Kotliar and Ruckenstein~\cite{PhysRevLett.57.1362}.
The holon-doublon correlator $C_{12}$ and double occupancy $\langle D \rangle$ as functions of $U$ are depicted in
Fig.~\ref{dope-vari-4site}(d) and (e), respectively, both of which decrease monotonically with $\delta$. $C_{12}$
in the PM state is smaller than that in the AFM state when $U<U_c$ ($8.9t$ for $\delta =0.02$), or otherwise
in the case of $U>U_c$, while the opposite is true for $\langle D \rangle$.
Furthermore, the ground state energy of the cluster slave-spin Hamiltonian (\ref{SSPIN-HAML})
$\langle H^S_{\text{4-site}}\rangle/2$ is plotted in  Fig.~\ref{dope-vari-4site}(f). There are three key points
needed to be addressed:
(i) According to Fig.~\ref{dope-vari-4site}(e), compared to the PM state, it is the decrease of the interaction
potential that makes $\langle H^S_{\text{4-site}}\rangle/2$ decline rapidly upon entering the AFM state.
(ii) For $U \lesssim U_{\text{Mott}} = 10t$, $\langle H^S_{\text{4-site}}\rangle/2$ increases monotonically with
doping, while it presents a nonmonotonic behavior  when $U \gtrsim U_{\text{Mott}}$.
(iii) In contrast to the PM state, $\langle D \rangle$ and $\langle H^S_{\text{4-site}}\rangle/2$ diminish
drastically as the AFM emerges, which can be understood through $\langle D _i\rangle=(1-\delta)/2-2\langle
M_i^2\rangle$ with $M_i=(n_{i\uparrow}-n_{i\downarrow})/2$.

We now discuss the difference of the energetics of the slave-spin Hamiltonian~\eqref{SSPIN-HAML} between the AFM
and PM states, as well as the electron momentum distribution and the hopping probability between the nearest and
next nearest neighbors.

\begin{figure}[htb!]
\centering{
\includegraphics[width=0.47\textwidth,scale=0.7]{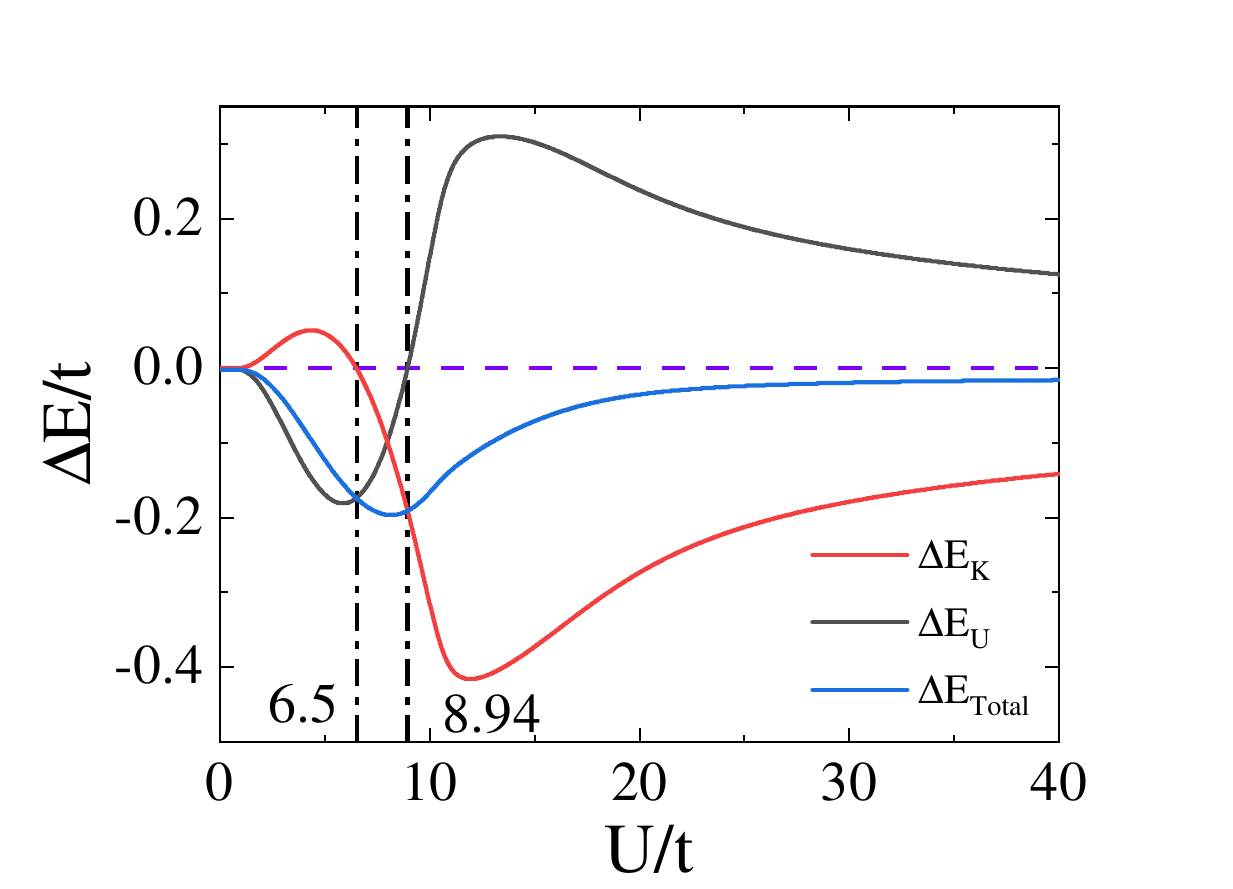}
}
\caption{The difference of the kinetic energy, interaction potential and the total energy of the cluster slave-spin
Hamiltonian between the AFM and PM  state as functions of $U$ at $\delta=0.02$ obtained by the four-site cluster.}
\label{EA-EP-DOPE002}
\end{figure}

The differences of the kinetic energy $E_{K}$, interaction potential $E_{U}$  and their summation
$E_{\text{Total}}$ of the approximate Hamiltonian (\ref{SSPIN-HAML}) between the AFM and PM states
($\Delta E=E^{(A)}-E^{(P)}$) as functions of $U$ are plotted in Fig.~\ref{EA-EP-DOPE002}.
We observe a region around $6.5t<U<8.94t$ separating the weak- and strong-coupling regimes in the AFM
state~\cite{PhysRevLett.124.117602,Mingpuqin2021,PhysRevB.94.085140,PhysRevB.94.195126,PhysRevLett.101.186403}.
When $U<6.5t$ or $ U>8.94t$, the AFM state in the system is governed by the interaction potential gain supporting
the Slater mechanism or the kinetic energy gain that favors the super-exchange mechanism,
respectively~\cite{Mingpuqin2021,PhysRevB.94.195126,PhysRevB.95.235109}. However, a sharp crossover is observed
at $U_{\text{Mott}}$ in the recent CDMFT calculation of the half-filed Hubbard model~\cite{PhysRevB.95.235109}.

The electron momentum distribution $n_d(\bm{k})$ calculated in Appendix~\ref{Fermi-Distribution} are plotted in
Fig.~\ref{NDK-2-4SITE}(a) and (b).
The overall feature is that the on-site interaction transfers the electrons within the Fermi surface outside and
the tendency is strengthened by the stronger interaction, which is consistent with the result of QMC
simulation~\cite{PhysRevB.80.075116}. However, we fail to reproduce the standard Fermi surface even in the
noninteracting limit, and $n_d(\bm{k})$ is larger than unity at $\bm{k}=(0,0)$ and negative at $\bm{k}=(\pi,\pi)$
in both two- and four-site cluster approximations, whereas an improvement can be seen when the cluster size is
enlarged. This kind of failure reflects an intrinsic flaw in all slave-variables approaches~\cite{PhysRevB.47.15192}.
\begin{figure}[htb!]
\centering{
\includegraphics[width=0.47\textwidth,scale=0.7]{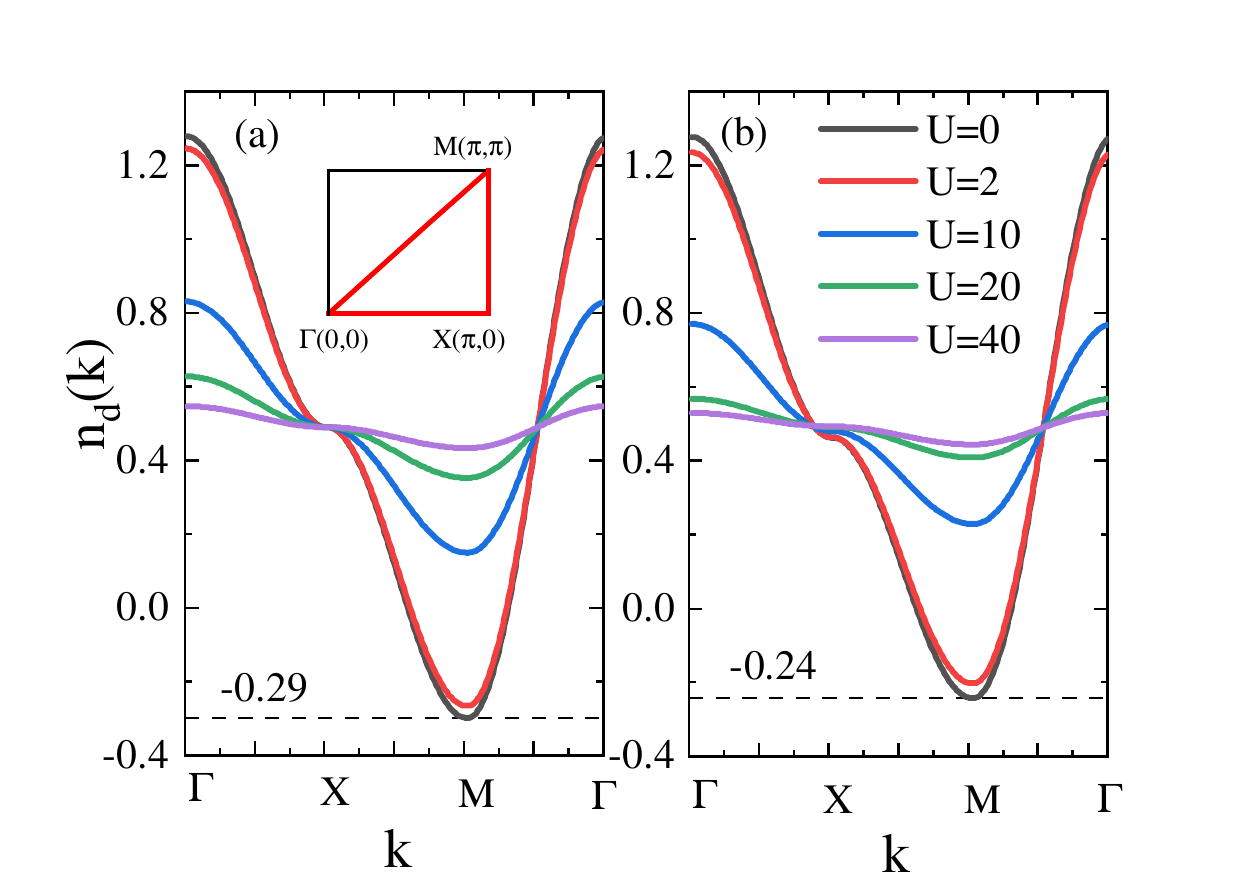}
}
\caption{(a), (b) The electron momentum distribution $n_d(\bm{k})$  versus $U$ at $\delta=0.02$ obtained by the
two and four-site cluster, respectively.}
\label{NDK-2-4SITE}
\end{figure}

\begin{figure}[htb!]
\centering{
\includegraphics[width=0.47\textwidth,scale=0.7]{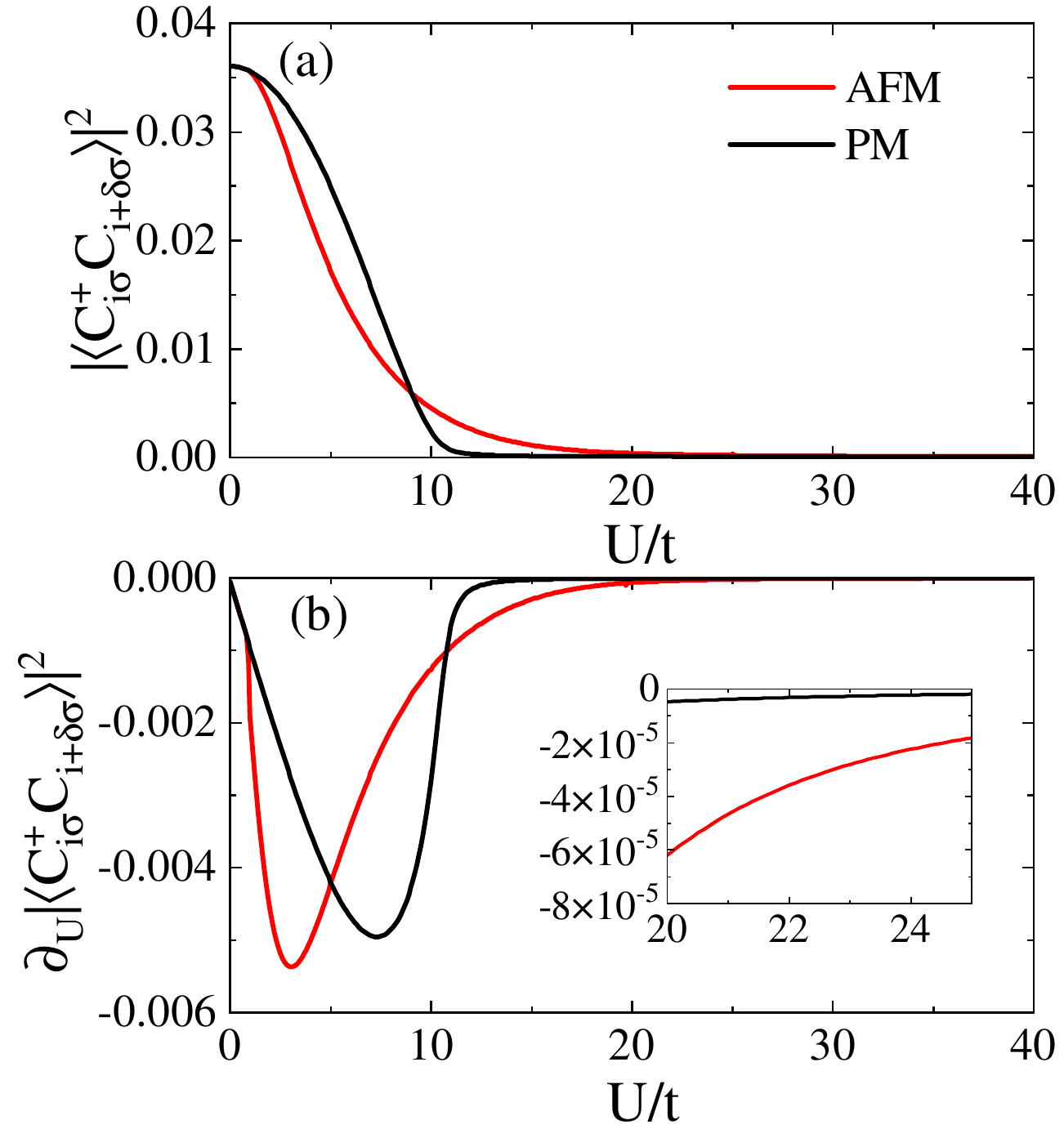}
}
\caption{(a) The hopping probability between the nearest neighbors $|\langle c^\dagger_{i\sigma}
c_{i+\delta\sigma}\rangle|^2$ and (b) its derivative as functions of $U$ at $\delta=0.02$ in the AFM (red)
and PM (black) states obtained by four-site cluster.}
\label{C1C2}
\end{figure}

The hopping probability between the nearest neighbor $|\langle c^\dagger_{i\sigma}c_{i+\delta\sigma}\rangle|^2\approx
Z^2 |\langle f_{i\sigma}f_{i+\delta\sigma}\rangle|^2$ and its derivative are plotted in Fig.~\ref{C1C2}(a) and (b).
$|\langle c^\dagger_{i\sigma}c_{i+\delta\sigma}\rangle|^2$ is smaller in the AFM state than in the PM state when
$U_{M}<U<U_c$, because the AFM state is driven by the interaction potential gain at small $U$
[see Fig.~\ref{EA-EP-DOPE002}] that suppresses the double occupancy, leading to the same effect on $|\langle
c^\dagger_{i\sigma}c_{i+\delta\sigma}\rangle|^2$ [compare Fig.~\ref{dope-vari-4site}(e) and Fig.~\ref{C1C2}(a)].
However, the opposite is true when $U>U_c$ because the AFM state is stabilized by the super-exchange mechanism
at large $U$. Accordingly, the gradient of $|\langle c^\dagger_{i\sigma}c_{i+\delta\sigma}\rangle|^2$ with respect
to $U$ has a minimum at $U= 3.0t,\; 7.4t$ in the AFM and PM states, respectively, signifying that the long range
AFM order suppresses the hopping probability between the nearest neighbors noticeably for small $U$.

\begin{figure}[htb!]
\centering{
\includegraphics[width=0.47\textwidth,scale=0.7]{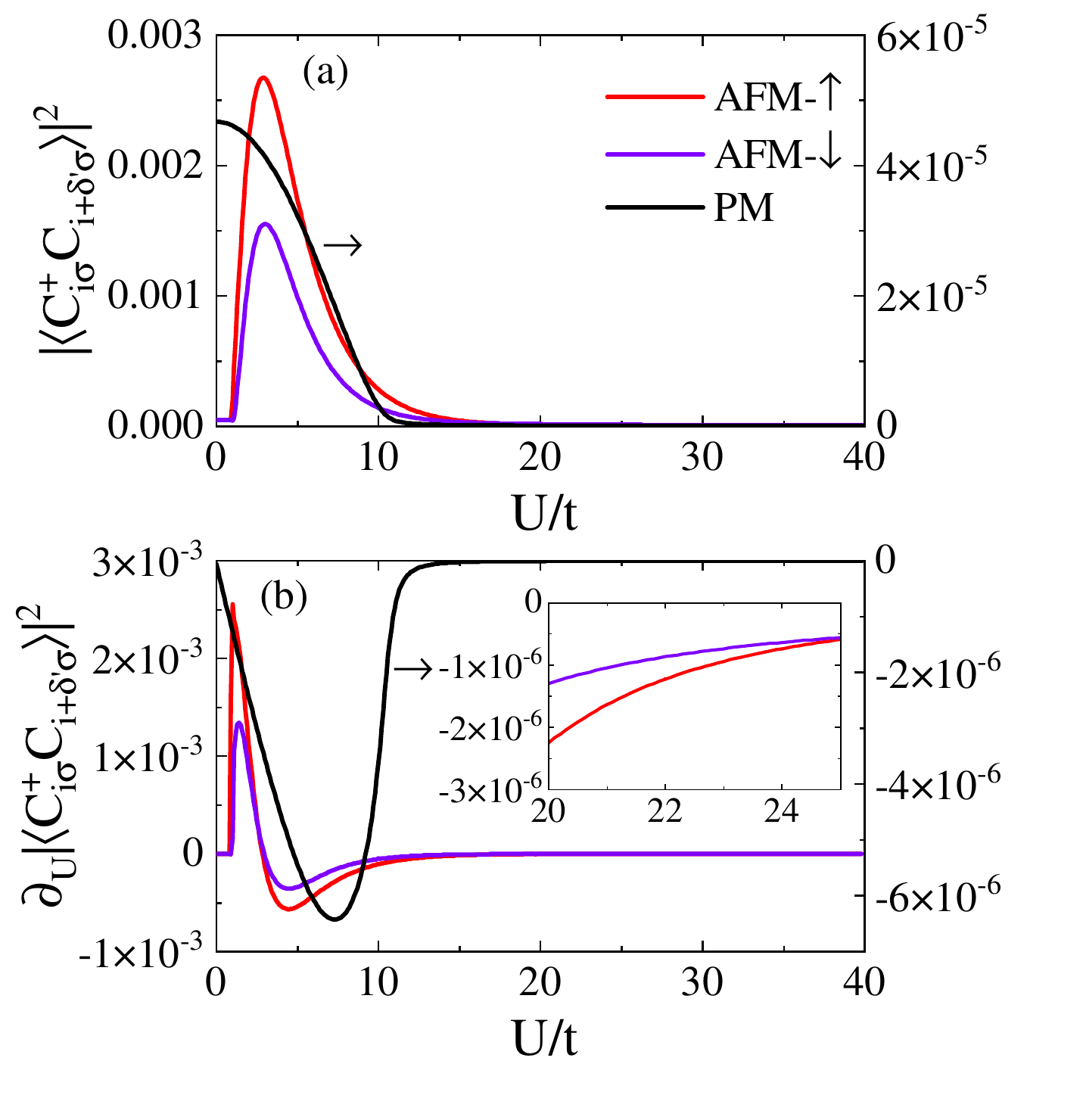}
}
\caption{(a) The hopping probability between the next nearest neighbors $|\langle c^\dagger_{i\sigma}
c_{i+\delta'\sigma}\rangle|^2$ and (b) its derivative as functions of $U$ at $\delta=0.02$ in the AFM (red and violet)
and PM (black) states obtained by four-site cluster.}
\label{C1C4}
\end{figure}

The hopping probability  between the next nearest neighbors $|\langle c^\dagger_{i\sigma}c_{i+\delta'\sigma}\rangle|^2\;
(i\in A) \approx  |\langle\tilde{z}_{A\sigma}\rangle|^4\langle f_{i\sigma}f_{i+\delta'\sigma}\rangle|^2$ and its
derivative are shown in Fig.~\ref{C1C4}(a) and (b).
In the AFM state, $|\langle c^\dagger_{i\sigma}c_{i+\delta'\sigma}\rangle|^2$  depends either on site $i$ or on spin
$\sigma$, and varies nonmonotonically --- remaining small when $U<U_{M}$, then zooming up at $U= U_{M}$, and vanishing
when $U> U_{\text{Mott}}$ after reaching its maximum at $U\sim 3t$. However, in the PM state, it decreases to zero
monotonically as $U$ increases.

\subsection{HALF-FILLED SYSTEM}\label{HALF-FILL}

By diagonalizing the spinon Hamitonian~\eqref{FERMI-HAML-K}, the self-consistent parameters
$n_{A/B\sigma}^f \equiv \langle f^\dagger_{A/B\sigma}f_{A/B\sigma}\rangle$ and $\epsilon^{\delta}_{\sigma} $
can be obtained
\begin{eqnarray}\label{EPS-F}
\epsilon_{\sigma}^{\delta}&=& \epsilon=\frac{1}{2N} \sum_{\bm{k}}
\frac{4t^2 Z \gamma_{\bm{k}}^2}{\sqrt{(4t Z \gamma_{\bm{k}})^2 + \Delta_{\sigma}^2}}
[ \theta(-E_{\bm{k}}^{+}) - \theta(-E_{\bm{k}}^{-})]\;, \nonumber \\
\end{eqnarray}
\begin{eqnarray}\label{NUM-F}
n_{A/B\sigma}^f
&=&\frac{1}{2N} \sum_{\bm{k}} \bigg[ \theta(-E_{\bm{k}}^{+}) + \theta(-E_{\bm{k}}^{-})\nonumber\\
&& \hspace{-2em} \mp \frac{\Delta_\sigma}{\sqrt{(4t Z \gamma_{\bm{k}})^2 + \Delta_{\sigma}^2}}
[\theta(-E_{\bm{k}}^{+}) - \theta(-E_{\bm{k}}^{-})] \bigg]\;,
\end{eqnarray}
where
\begin{equation}
E_{\bm{k}}^{\pm}= - \mu^{\text{eff}} \pm \sqrt{(4t Z \gamma_{\bm{k}})^2 + \Delta_{\sigma}^2}\;.
\end{equation}
At half-filling, the particle-hole symmetry ensures that $\mu^{\text{eff}}=0$ and $E_{\bm{k}}^{+}>0$. With the density of state \eqref{density},  Eqs.~\eqref{EPS-F} and \eqref{NUM-F} can be rewritten as
\begin{eqnarray}
\epsilon &=& - t\lambda \int_0^1 d\gamma D(\gamma)\,\gamma^2
\frac{1}{\sqrt{(\lambda\gamma)^2 +1}}\nonumber \\
&=& - t\lambda I_\epsilon(\lambda)\;, \\
n_{A/B\sigma}^f
&=& \frac{1}{2}\mp  \,{\rm sgn}({\Delta}_\sigma) \int_0^1 d\gamma D(\gamma)\, \frac{1}{\sqrt{(\lambda\gamma)^2 +1}} \nonumber \\
&=& \frac{1}{2}\mp\,{\rm sgn}({\Delta}_\sigma)\, I_f(\lambda)=\frac{1}{2}+ \langle S_{A/B\sigma}^z \rangle\;.
\end{eqnarray}
with $\lambda = 4tZ/\Delta_{\text{AFM}}$.

\begin{figure}[htb!]
\centering{
\includegraphics[
scale=0.7]{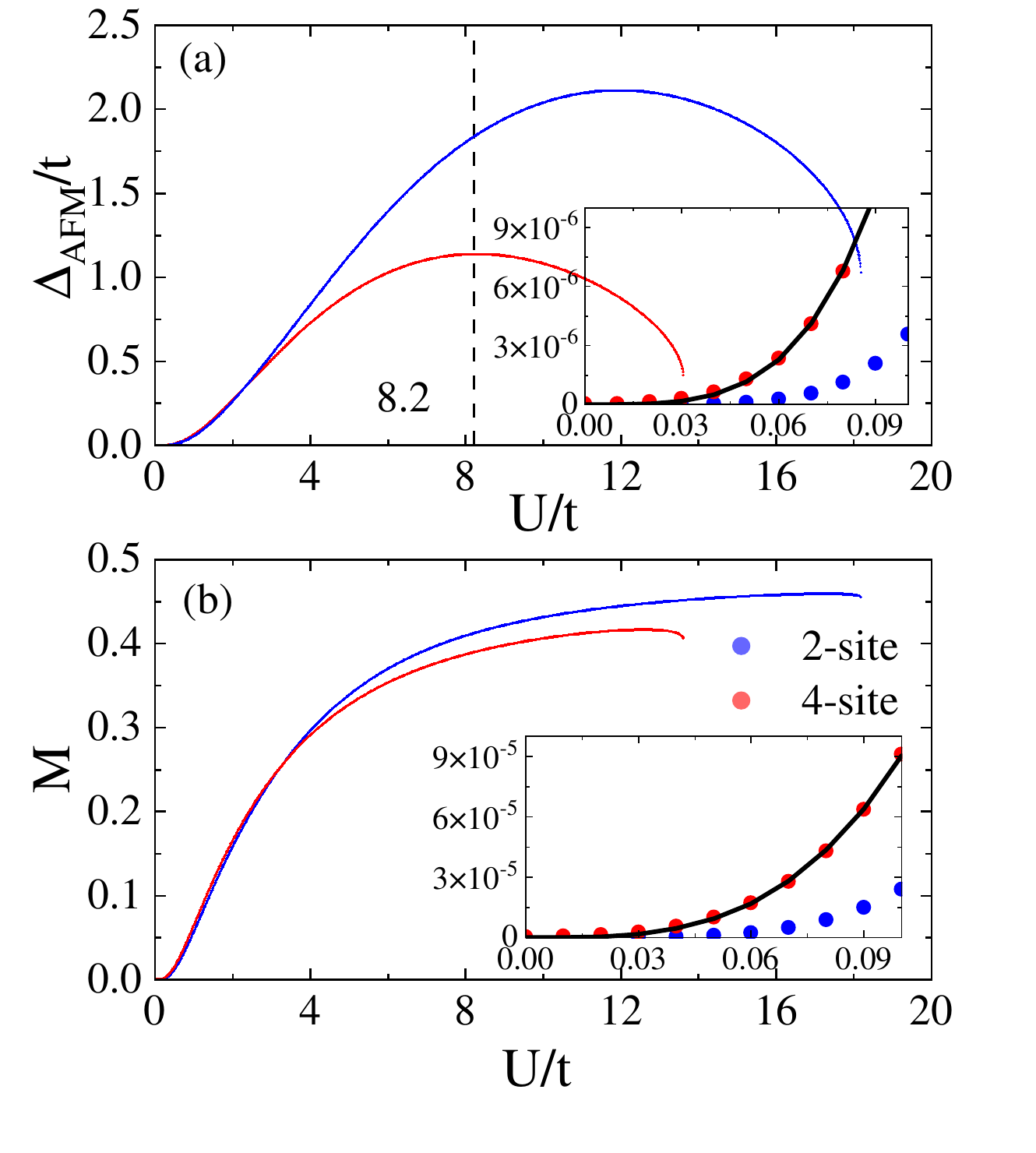}
}
\caption{(a) the AFM energy gap $\Delta_{\text{AFM}}/t$  where the inset shows the same data at small $U$ together
with the fitting data (black line)
and (b) the staggered magnetization $M$ where the inset shows the same data at small $U$ together with that calculated by
Eq.~(\ref{STAG-M-SSPIN}) (black line).}
\label{M-DELTA-DOPE00}
\end{figure}

To establish the relation between $M$ and $\Delta_{\text{AFM}}$  in the small $U$ limit, we expand asymptotically the
integrals $I_\epsilon(\lambda)$ and $ I_f(\lambda) $ as $\lambda\to\infty$~\cite{NBRAH}:
\begin{eqnarray}
\epsilon &\sim& -t \bigg[ \frac{2}{\pi^2}
- \frac{\tilde{\Delta}^3 (\ln \tilde{\Delta})^2 }{2\pi^2}
- \Big( \frac{1}{2\pi^2} - \frac{3\ln 2}{\pi^2} \Big) \tilde{\Delta}^2 \ln \tilde{\Delta}\nonumber\\
&& \hspace{2em} + \Big( \frac{1}{4\pi^2} + \frac{3 \ln 2}{2\pi^2} - \frac{9 (\ln 2)^2}{2\pi^2} \Big) \tilde{\Delta}^2
\bigg]\;,\label{EPSLON-SSPIN}
\end{eqnarray}
\begin{eqnarray}
\langle S_{A/B\sigma}^z \rangle &\sim& (-/+) \,{\rm sgn}({\Delta}_\sigma)\, \bigg[ \frac{1}{\pi^2} \tilde{\Delta} (\ln
\tilde{\Delta})^2 \nonumber\\
&& \hspace{2em} - \frac{6\ln 2}{\pi^2} \tilde{\Delta} \ln \tilde{\Delta}
+ \frac{9(\ln 2)^2}{\pi^2} \tilde{\Delta} \bigg] \;,\label{Delta-SSPIN}
\end{eqnarray}
with $ \tilde{\Delta}=\lambda^{-1}=\Delta_{\text{AFM}}/(4tZ)$, resulting in
\begin{eqnarray}\label{STAG-M-SSPIN}
M &=& \tfrac{1}{2}|\langle S_{A\sigma}^z \rangle-\langle S_{B\sigma}^z \rangle|\nonumber\\
&\sim& \frac{1}{\pi^2} \tilde{\Delta} (\ln \tilde{\Delta})^2 - \frac{6\ln 2}{\pi^2} \tilde{\Delta} \ln \tilde{\Delta}
+ \frac{9(\ln 2)^2}{\pi^2} \tilde{\Delta} \;.
\end{eqnarray}
Therefore, the simple HF relation of $\Delta_{\text{AFM}} = UM$ (Ref.~\onlinecite{PhysRevB.31.4403}) needs to
be corrected. By adopting the $U$ dependent form of the AFM gap
\begin{equation}\label{Delta-M-DLT0}
\Delta_{\text{AFM}} = A\sqrt{U}e^{-B\sqrt{\frac{1}{U}}}\;,
\end{equation}
which has been verified by the QMC method~\cite{PhysRevB.31.4403}, we fit our self-consistent data to obtain
\begin{eqnarray}
\Big\{ \begin{array}{l} A=0.02086\pm 0.00418\;, \\ B=1.90604\pm0.06189 \;, \end{array}
\end{eqnarray}
as shown in Fig.~\ref{M-DELTA-DOPE00} (a). Then, the result is substituted into Eq.~\eqref{STAG-M-SSPIN} to recover
perfectly the data of $M$ shown Fig.~\ref{M-DELTA-DOPE00} (b).

Moreover, the AFM gap $\Delta_{\text{AFM}}$, starting from an infinitesimal $U$, should also present in the spectrum
function of the electron~\cite{PhysRevB.94.085140}.
 In the slave-spin method, the spectrum function of the electron Green's function can be expressed as\cite{PhysRevB.86.045128}
\begin{equation}
A(\bm{q},\omega)=\int_{\bm{k}}\int_0^\omega d\Omega A_{\text{s-spin}}(\bm{q}-\bm{k},\omega-\Omega)
A_{\text{fermion}}(\bm{k},\Omega)\;.
\end{equation}
Assuming a gap $\Delta_{\text{AFM}}$ in the fermionic spinon sector of Hamiltonian~\eqref{Hubbard model},
$A_{\text{fermion}}(\bm{k},\Omega<\Delta_{\text{AFM}})=0$, it is easy to understand that the electron spectrum function
$A(\bm{q},\omega)$ will vanish when $\omega$ is smaller than $\Delta_{\text{AFM}}$.

\begin{figure*}[htb!]
\centering{
\includegraphics[
scale=0.48]{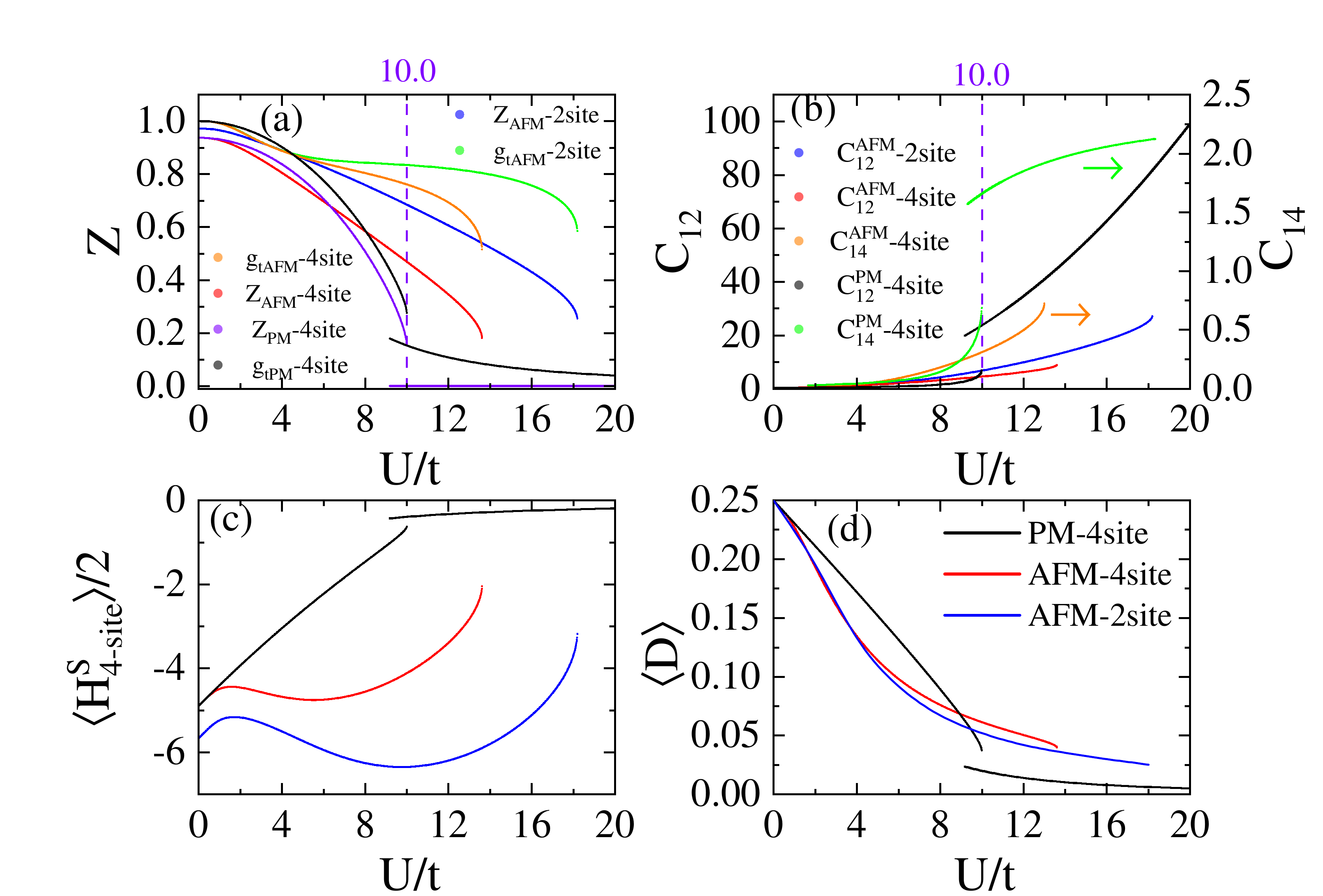}
}
\caption{(a) The quasiparticle weight $Z$ and the generalized Gutzwiller factor $g_t$,  (b) the holon-doublon correlators
between the nearest neighbors $C_{12}$ and the next nearest neighbors $C_{14}$, (c) the expectation value of the cluster
slave-spin Hamiltonian,  and (d) the double occupancy $\langle D\rangle$ in the AFM state as functions of  $U$ obtained
by the 2/4-site clusters (blue, red), respectively, and those in the PM state obtained by the 4-site cluster (black).}
\label{dope00-2-4site}
\end{figure*}

\begin{figure}[htb!]
\centering{
\includegraphics[width=0.47\textwidth,scale=0.7]{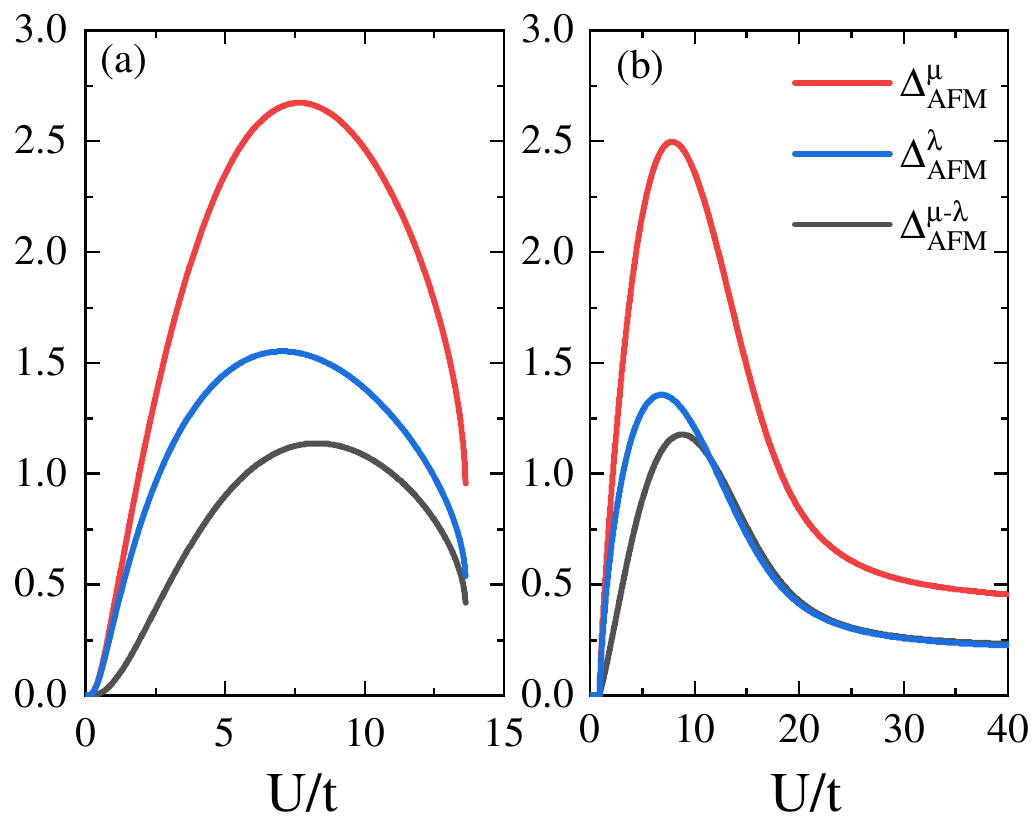}
}
\caption{The AFM gap $\Delta_{\text{AFM}}/t$ (black), $\Delta^{\mu}_{\text{AFM}}/t$ (red), $\Delta^{\lambda}_{\text{AFM}}/t$
(blue) as functions of  $U$ obtained from the four-site cluster at (a) $\delta=0.0$ and (b) $\delta=0.02$.}
\label{DLT-DOPE-00-002}
\end{figure}

Additionally, $Z$, $C_{12/14}$, $\langle H_{\text{4-site}}^{S}\rangle/2$, and $\langle D \rangle$ as functions of $U$
are plotted in Fig.~\ref{dope00-2-4site}.
The overall features are similar to those with $\delta=0.02$ [cf. (Fig.~\ref{SSPINdope002}) for comparison] except two critical
distinctions. First, in the PM state, the first-order Mott transition occurs at $U_{\text{Mott}} =10t$, characterized by the
discontinuity and hysteresis behavior of these quantities~\cite{RevModPhys.68.13,PhysRevLett.101.186403,PhysRevLett.110.096402,
JPSJ.80.084705}. Compared to the AFM's $U_{c}$=8.2t shown in Fig.~\ref{M-DELTA-DOPE00}(a), we find that the AFM correlations
significantly reduce the coupling strength which separates the weak- and strong-coupling regimes. This is consistent with the
CDMFT calculation~\cite{PhysRevLett.101.186403}. Second, both the AFM solutions at large $U$ obtained from the 2/4-site cluster
approximations are absent, and the quasi-particle residue drops abruptly when $U$ approaches the critical coupling strength
where the AFM solution happens to disappear, which may imply a transition between the Hubbard model with $Z\neq0$ and Heisenberg
model with $\Delta_{\text{AFM}}=0$, i.e., $Z=0$.
To justify this argument, notice the contribution to $\Delta_{\text{AFM}}$ comes from two parts [see Eq.~(\ref{defDLTAFM})]:
the differences of the effective chemical potentials [$\Delta_{\text{AFM}}^{\mu}=\tfrac{1}{2} (\tilde{\mu}_{A\sigma} -
\tilde{\mu}_{B\sigma} )$] and the lagrange multipliers
[$ \Delta_{\text{AFM}}^{\lambda}=\tfrac{1}{2} (\lambda_{A\sigma} - \lambda_{B\sigma})$] between sublattice A and B,
i.e., $\Delta_{\text{AFM}}=\Delta^{\mu-\lambda}_{\text{AFM}}=\Delta^{\mu}_{\text{AFM}}-\Delta^{\lambda}_{\text{AFM}}$, all
of which as functions of $U$ are plotted in Fig.~\ref{DLT-DOPE-00-002}. For comparison, those at $\delta=0.02$ are presented
too. One finds the main contribution to $\Delta_{\text{AFM}}$ is from $\Delta_{\text{AFM}}^{\mu}$~\cite{PhysRevB.96.115114},
which is proportional to $Z$. From Fig.~\ref{DLT-DOPE-00-002}(a), both $\Delta_{\text{AFM}}^{\mu}$ and
$\Delta_{\text{AFM}}^{\lambda}$ approach zero as $Z$ vanishes, resulting in the disappearance of $\Delta_{\text{AFM}}$ at
$U\sim13.6t$, which may be connected to the gapless
magnetic excitation of the Heisenberg model.

\subsection{PHASE DIAGRAM}\label{PHASES}

\begin{figure}[htb!]
\centering{
\includegraphics[width=0.47\textwidth,scale=0.7]{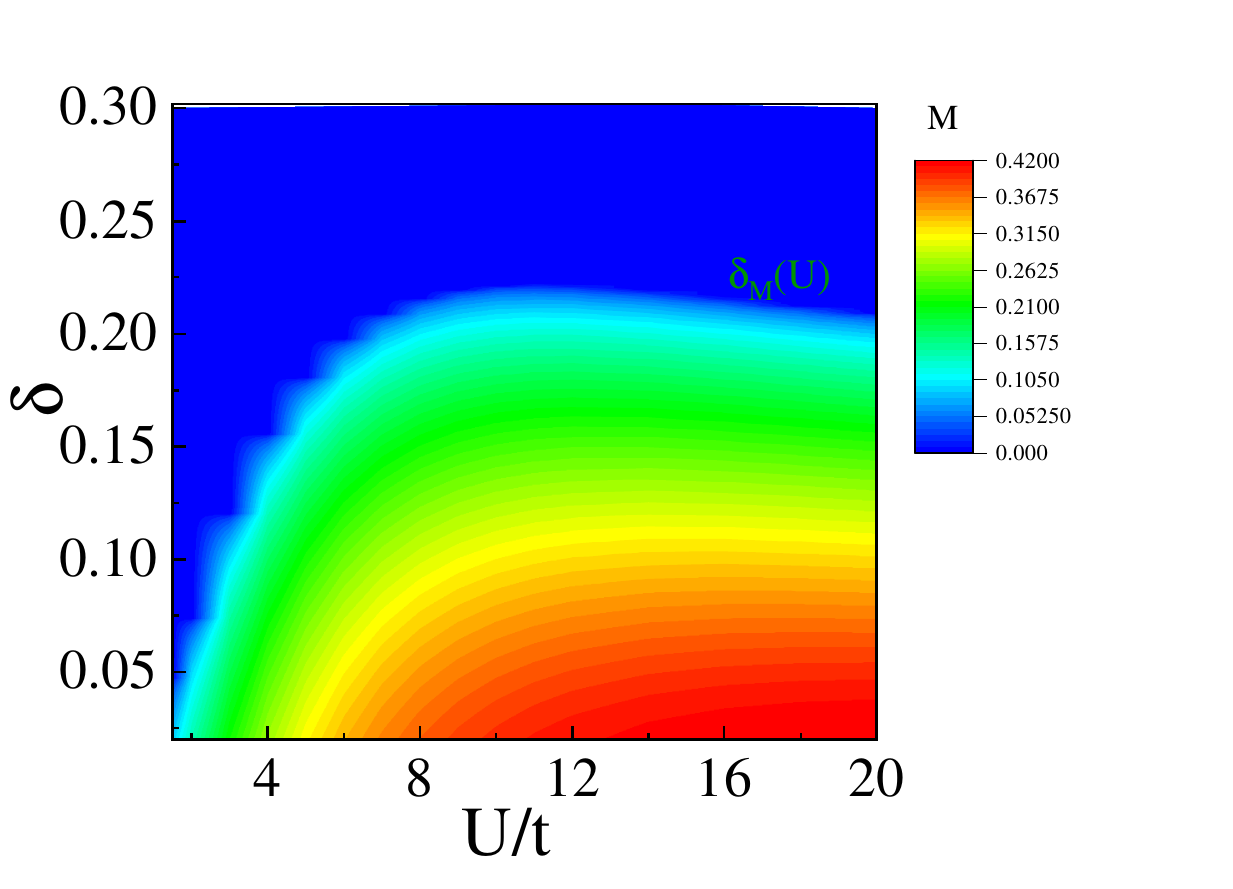}
}
\caption{The staggered magnetization $M$  as functions of $U$ and $\delta$ in the AFM states obtained by the four-site cluster.}
\label{M-DOPE-U}
\end{figure}

\begin{figure}[htb!]
\centering{
\includegraphics[width=0.47\textwidth,scale=0.7]{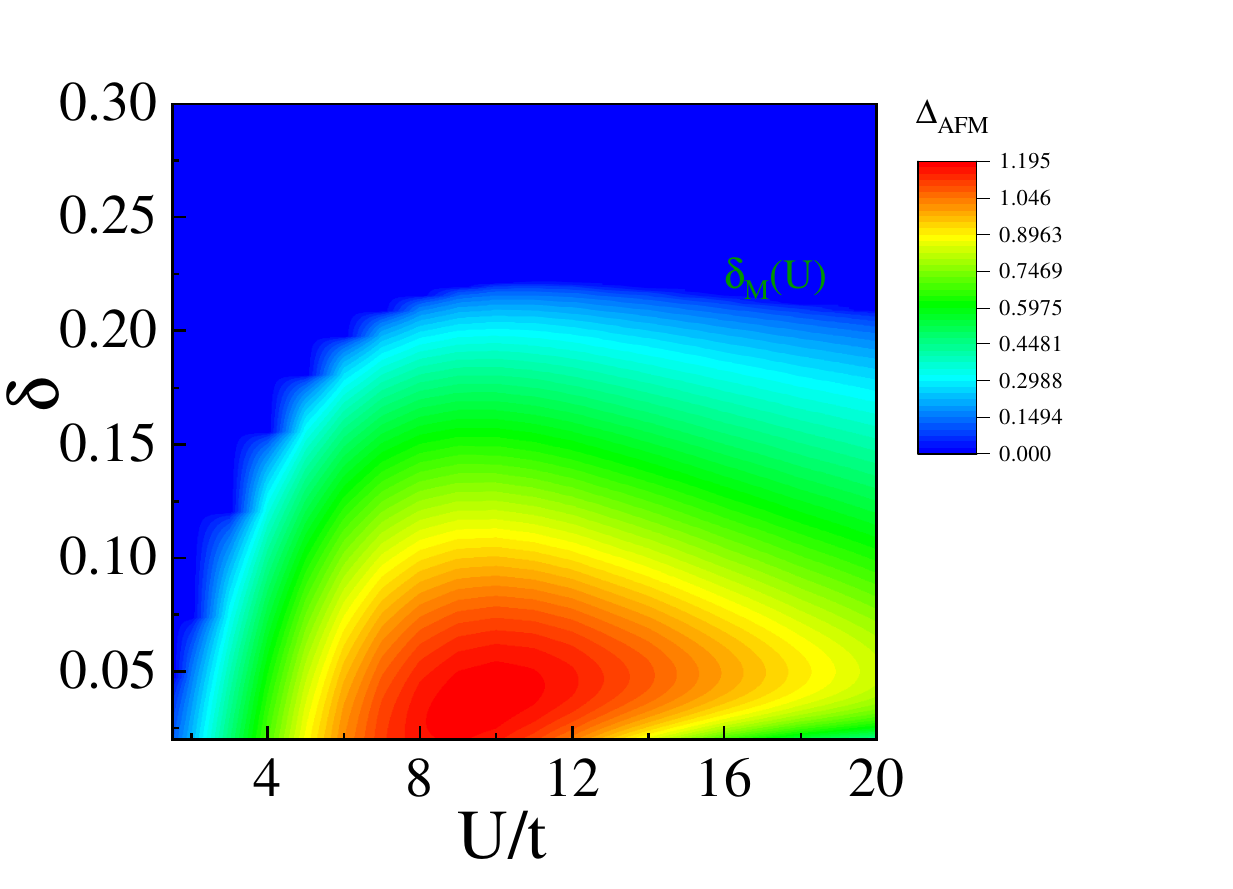}
}
\caption{The AFM gap $\Delta_{\text{AFM}}$  as functions of $U$ and $\delta$  in the AFM states obtained by the four-site cluster.}
\label{DELTA-DOPE-U}
\end{figure}

\begin{figure}[htb!]
\centering{
\includegraphics[width=0.47\textwidth,scale=0.7]{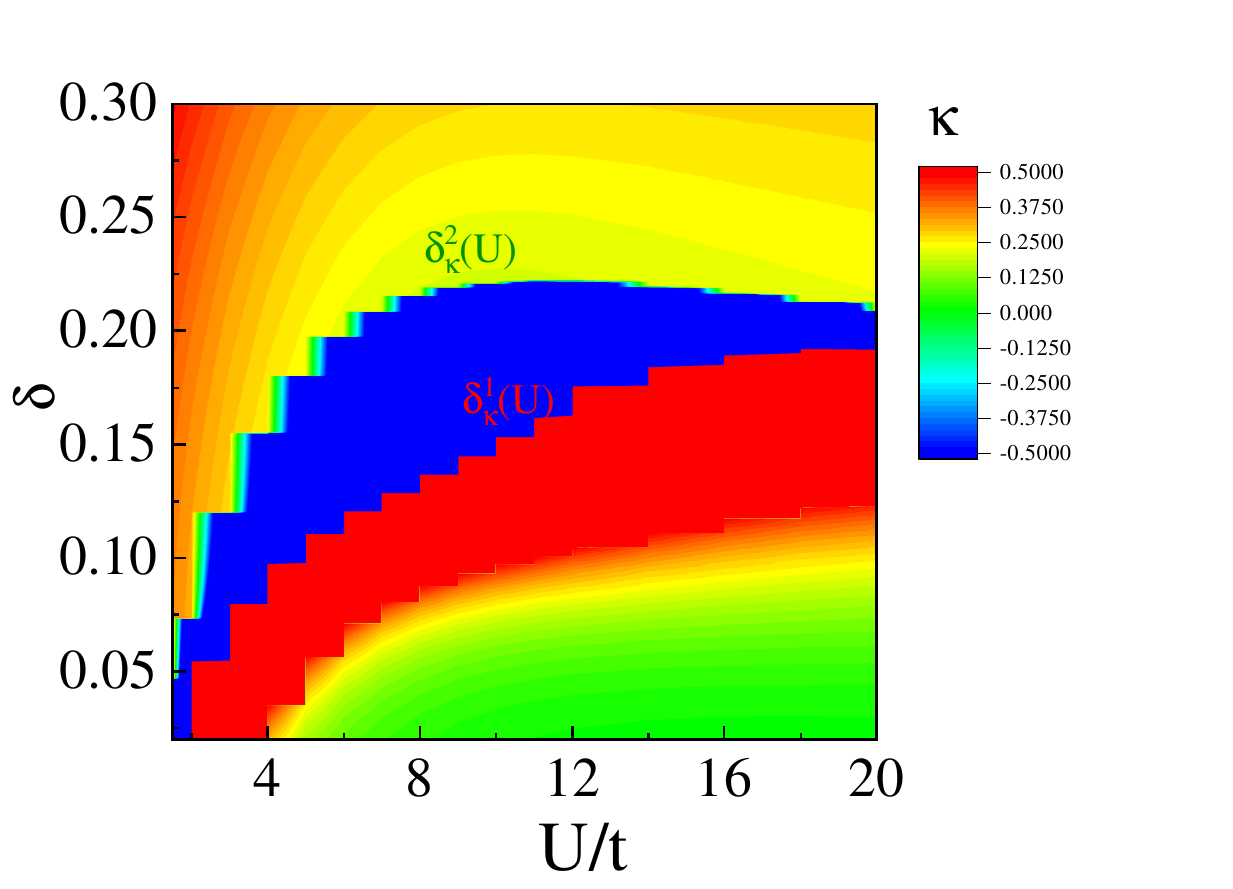}
}
\caption{The compressibility $\kappa$ as functions of $U$ and $\delta$ in the AFM states obtained by the
four-site cluster. $\kappa$ in the red and blue region are greater than 0.5 and less than -0.5, respectively.}
\label{KAPPA-DOPE-U}
\end{figure}

\begin{figure}[htb!]
\centering{
\includegraphics[width=0.47\textwidth,scale=0.7]{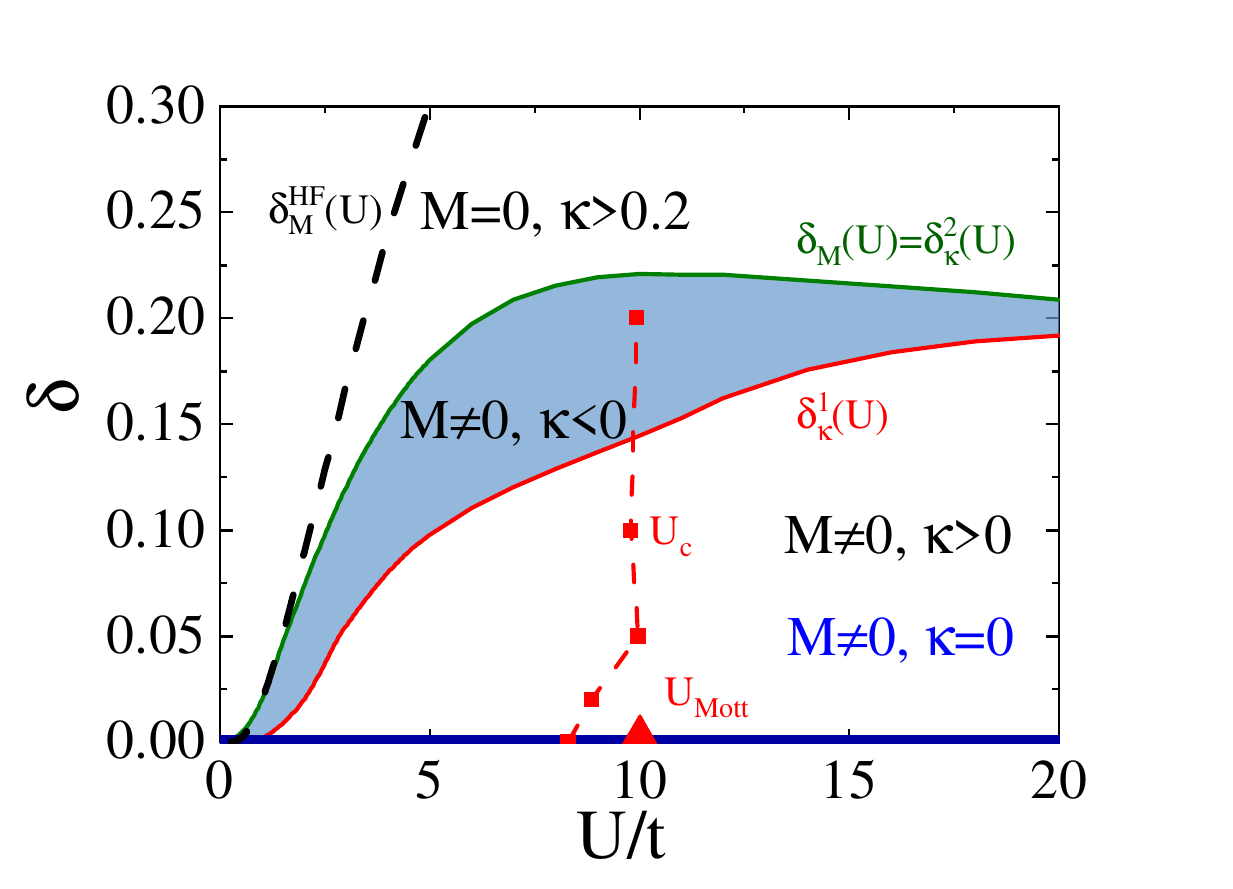}
}
\caption{The phase diagram of the Hubbard model in the $U$-$\delta$ plane obtained by the four-site cluster,
where exist the crossover $U_{c}$ (red dashed line) dividing the weak- and strong-coupling regimes, at which the
$\Delta_{\text{AFM}}$ is maximized, and the AFM
for $U< U_{c}$ and $U> U_{c}$ is advocated by the interaction potential and kinetic energy gain, respectively.
The $U_{\text{Mott}}$ for the Mott transition in the PM state is marked by the red triangle.
The half-filling case is highlighted by the heavy blue line, in which the system is an AFM insulator with $\kappa=0$.
Here the black dashed line represents the boundary $\delta_{\text{MF}}(U)$
between the AFM and PM phases within the HF method.
}
\label{DOPE-U-PHASES}
\end{figure}

In the following, we combine the results of the staggered magnetization $M$ (Fig.~\ref{M-DOPE-U}), the AFM gap
$\Delta_{\text{AFM}}$ (Fig.~\ref{DELTA-DOPE-U}), and the compressibility $\kappa$ (Fig.~\ref{KAPPA-DOPE-U}) to
present an overall phase diagram for the Hubbard model (Fig.~\ref{DOPE-U-PHASES}).

The staggered magnetization $M$ with $\delta$ and $U$ being its parameters is plotted in Fig.~\ref{M-DOPE-U}, where the
phase boundary between the AFM and PM states is denoted by $\delta_{M}(U)$.
First, the staggered magnetization always saturates at a certain value when $U> U_{\text{Mott}}$, while decreases
monotonically with $\delta$.
Second, marked by the dense contours around $\delta_{M}(U)$, the staggered magnetization decreases continuously
to zero as $\delta$ approaches $\delta_{M}(U)$, signifying the second-order transition between the AFM and the PM phases.
Third, $\delta_{M}(U)$ shows a nonmonotonic behavior with $U$ (Ref.~\onlinecite{PhysRevB.96.241109}), which is consistent with the
re-entrance of $M$ with $U$ at $\delta= 0.2$, as shown in Fig.~\ref{dope-vari-4site}(c).

The AFM gap $\Delta_{\text{AFM}}$ in the same parameter space is plotted in Fig.~\ref{DELTA-DOPE-U},
where the previous boundary $\delta_{M}(U)$ still holds for $\Delta_{\text{AFM}}$.
When $U>U_{\text{Mott}}$, the maximum of $\Delta_{\text{AFM}}$ occurs at $\delta\approx 0.05$, leading to the
interesting vertical re-entrance behavior as $\delta$ increases,  which reflects that the AFM gap at half-filling
vanishes in the large $U$ limit.

\begin{figure}[htb!]
\centering{
\includegraphics[width=0.47\textwidth,scale=0.7]{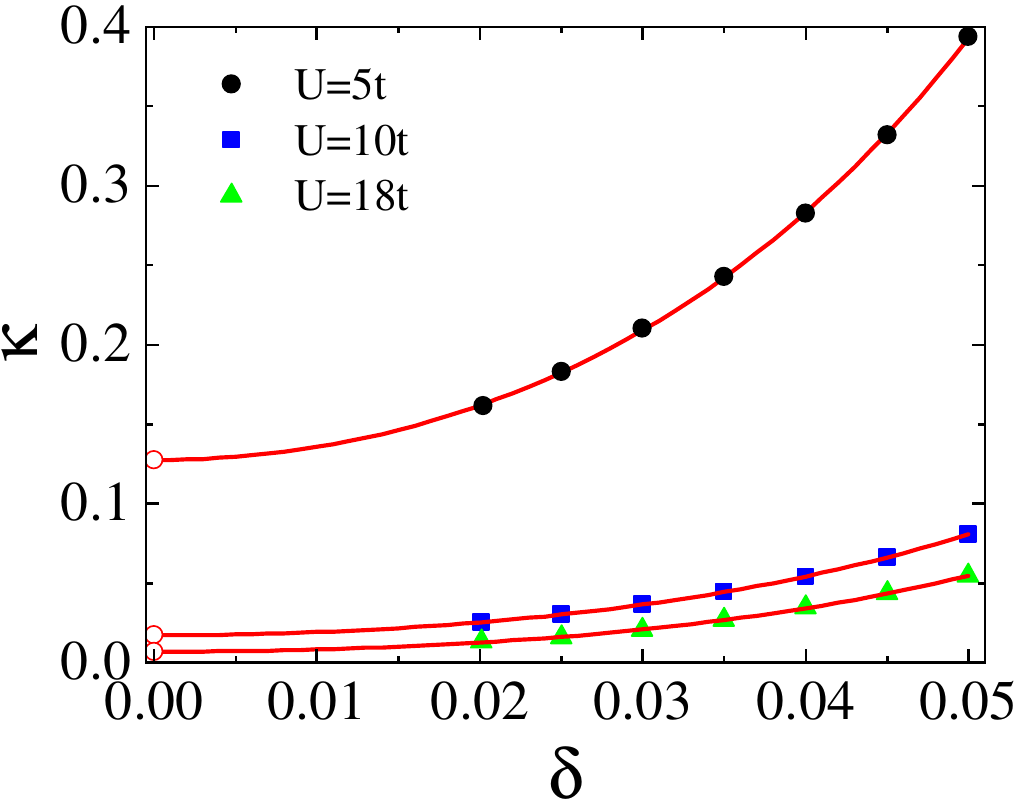}
}
\caption{The compressibility $\kappa$ as functions of $\delta$ in the AFM state at $U= 5t$, $10t$, and $18t$
obtained by four-site cluster. Red lines are the extrapolated curves from the corresponding self-consistent data
using $\kappa_0 + \alpha \delta^2 + \beta \delta^4$.
The red circles at $\delta=0$ denotes the discontinuity of $\kappa =0$ at $\delta=0$.}
\label{KAPPA-DOPE}
\end{figure}

Figure~\ref{KAPPA-DOPE-U} shows the results for the compressibility $\kappa= n^{-2}\partial n/\partial\mu$.
The staircase in this figure obviously is an artifact because of discrete $U$'s adopted to calculate $\kappa$,
i.e., $\Delta U=t$
when $U\le 12t$ and $\Delta U=2t$ when $12t<U\le20t$, which can only be eliminated in the $\Delta U\to 0$ limit.
As shown in Fig.~\ref{KAPPA-DOPE-U}, there exist two phase boundaries delineated by the midpoints of these steps:
(i) $\delta_{\kappa}^{1}(U)$, between the regions with positive (red) and negative (blue) compressibility.
(ii) $\delta_{\kappa}^{2}(U)$, from the region with $\kappa<-0.5$ (blue) to $0.2<\kappa<0.5$ (yellow).
As shown in Fig.~\ref{KAPPA-DOPE}, $\kappa=0$ at half-filling is disconnected from that at an infinitesimal
doping~\cite{PhysRevB.96.241109}, signifying that the half-filled system in the AFM state is an insulator while
the doped one a metal.
By Fig.~\ref{M-DOPE-U} and Fig.~\ref{KAPPA-DOPE-U},
the systems with $\delta_{\kappa}^{1}(U)<\delta<\delta_{\kappa}^{2}(U)$ are AFM
metals with negative compressibility, meaning the uniform AFM configuration is not the actual ground state, whereas
an inhomogeneous phase with $M\neq0$ could be the
alternaltive~\cite{Zheng2016,PhysRevB.94.195126,PhysRevB.96.081117,PhysRevB.91.241116,PhysRevB.78.165101}.
In addition, the space between $\delta_{\kappa}^{1}(U)$ and $\delta_{\kappa}^{2}(U)$ shows a broadening-narrowing
feature as $U$ increases, and they may merge into one as $U\gg 20t$, implying that this kind of phase
separation can not be observed in the large $U$ limit~\cite{PhysRevB.78.165101,PhysRevB.91.241116,PhysRevB.94.195126}.
Moreover, $\delta_{\kappa}^{2}(U)$  is identical to $\delta_{M}(U)$, thus, $\delta_{M}(U)$ is the boundary separating
the AFM metal phase with $\kappa<0$ and the PM metal phase.

By aggregating Figs.~\ref{M-DOPE-U}, \ref{DELTA-DOPE-U} and \ref{KAPPA-DOPE-U}, we acquire an overall phase diagram,
Fig.~\ref{DOPE-U-PHASES}, of the Hubbard model in the $U$-$\delta$ plane. The boundary $\delta_{\text{HF}}(U)$
between the AFM and PM phases within the HF theory is plotted as the dashed line. As expected, the consistency between
$\delta_{\text{HF}}(U)$ and $\delta_{M}(U)$ only lies in the small $U$ region, which is also the case of $M$ in
Fig.~\ref{SSPINdope002}(c). There exists one
crossover~\cite{PhysRevB.94.195126,PhysRevB.95.235109,PhysRevB.96.115114,PhysRevLett.124.117602} $U_{c}$,
separating the weak- and strong-coupling regions, and three transitions in the $U$-$\delta$ plane:
(i) between the AFM insulator at $\delta=0$, marked by the heavy blue line, and the AFM metal for $\delta>0$;
(ii) $\delta_{\kappa}^{1}(U)$, separating the AFM phases with the positive and negative compressibility, which may
not exist after considering the inhomogeneous phases;
 (iii) $\delta_{\kappa}^{2}(U)$, from the AFM metal with negative compressibility to the PM metal.
 Thus, we find phase separation at small and intermediate coupling strengths, which is in agreement with the
 auxiliary-field QMC~\cite{PhysRevB.78.165101,PhysRevB.91.241116} and
 variational~\cite{PhysRevB.94.195126} studies. However, the phase separation occurs at
 intermediate doping levels in our work but at small dopings in these QMC
 studies. This discrepancy most likely results from the fact that the cluster slave-spin mean-field theory
 exaggerates the system's tendency towards a uniform AFM state, which may
 be remedied by enlarging the cluster size and strictly dealing with the
 constraint $S^{z}_{\alpha}=f^{\dagger}_{\alpha}f_{\alpha}-\tfrac{1}{2}$ locally within the cluster. 
 It should be mentioned that $U_{c}$'s at all dopings are close to the $U_{\text{Mott}}$ of $\delta=0$,
 manifesting that the physics in  the AFM state are governed by the underlying Mott transition in the half-filled PM
 state~\cite{PhysRevB.96.241109,PhysRevB.95.235109,PhysRevB.94.195126}. The reason for the close relationship between the
 crossover mentioned above and the Mott transition in the PM state is that both phenomena are driven by the
 competition between the kinetic energy and interaction potential of the Hubbard model.

\section{CONCLUSION}\label{conclude}

In summary, the cluster slave-spin method has been employed to investigate systematically the ground state properties
of the single-band 2D Hubbard model on a square lattice in the parameter space of $U$ and $\delta$. We substantiated that the system presents
a broad crossover between the weak- and strong-coupling regimes: (i) The $\Delta_{\text{AFM}}$
increases monotonically with $U$ in the weak-coupling regime, while decreases for large couplings~\cite{PhysRevB.96.115114}.
(ii) The AFM in the weak-coupling regime is stabilized by the interaction potential gain, while that for large $U$ is supported by
the kinetic energy gain~\cite{Mingpuqin2021,PhysRevB.94.195126,PhysRevB.95.235109}. In Fig.~\ref{DOPE-U-PHASES},
the $U_{c}$ marked by the red dashed line which is close to $U_{\text{Mott}} = 10t$, signifying that the underlying
Mott transition in the half-filled PM state dominates the properties of the AFM states by changing into a
crossover~\cite{PhysRevB.94.195126,PhysRevB.95.235109,PhysRevB.96.241109} between the weak- and strong-coupling regimes.
It is worthy mentioning that $U_{c}$ at half-filling is smaller than $U_{\text{Mott}}$ because of
long range AFM correlations~\cite{PhysRevLett.101.186403}.

For the half-filled system, we analytically calculated in Eq.~\eqref{STAG-M-SSPIN} the relation between $M$ and
 $\Delta_{\text{AFM}}$ in the small $U$ limit, implying that the one from the HF method needs to be
 improved to include quantum fluctuations. After fitting the AFM gap of the cluster slave-spin data using the
 formula from the QMC simulation~\cite{PhysRevB.31.4403}, and calculating the dependence of $M$ upon $U$ by
 Eq.~\eqref{STAG-M-SSPIN}, we find the result is in a good agreement with the numerical data
 as shown in Fig.~\ref{M-DELTA-DOPE00}.

The $U$-$\delta$ phase diagram was discussed in Sec.~\ref{PHASES}, from which we extracted four regimes: AFM insulator
at $\delta=0$, AFM metal with $\kappa>0$ and $\kappa<0$, and the PM metal. The second-order transition occurs when the
system transits from the AFM metal with $\kappa<0$ to the PM metal phase, whereas the existence of the transition from
the AFM metal phase with $\kappa>0$  to that with $\kappa<0$ needs to be proven further by taking the inhomogeneous
states into account. Moreover, as shown in Figs.~\ref{DELTA-DOPE-U} and \ref{DOPE-U-PHASES}, the crossover,
separating the weak- and strong- coupling regimes in the AFM state, always locates around $U_{\text{Mott}}$,
and whether this property is specifically associated with the geometry  of the square lattice is worthy of
subsequent investigations.

According to Figs.~\ref{SSPINdope002} and \ref{dope00-2-4site}, the difference between the two- and four-site
cluster results is rather quantitative than qualitative. We believe that the four-site
cluster approximation is sufficient to capture the physics because its geometry obeys the same lattice
symmetry as the original square lattice, which is very important to extract the reasonable solutions of the system. This
conclusion can be reached in our ongoing studies on the honeycomb lattice.

\begin{acknowledgments}
This work was supported by NSFC (Nos. 11774033 and 11974049) and Beijing Natural Science Foundation (No.1192011).
Q. Chen would like to thank the initial funding from NSFC (Grant No. 11504023).
We acknowledge the support of HSCC of Beijing Normal University, and some numerical simulations
in this work were performed on
Tianhe in Beijing Computational Science Research Center.
\end{acknowledgments}

\appendix

\section{RELATIONSHIP BETWEEN THE HOLON-DOUBLON CORRELATOR AND THE MOTT TRANSITION }\label{Correlator}

The holon-doublon correlator~\cite{PhysRevB.96.115114} is defined as
\begin{equation}\label{ho_do_co}
C_{ij}=\frac{\langle N_{i}D_{j}\rangle-\langle N_{i}\rangle \langle D_{j}\rangle}{\langle N_{i}\rangle \langle
D_{j}\rangle}\;,
\end{equation}
where
\begin{equation}
N_{i}=(1- c^{\dagger}_{i\sigma} c_{i\sigma}) (1- c^{\dagger}_{i\bar \sigma} c_{i\bar\sigma}), \hspace{0.5em}
D_{j}= c^{\dagger}_{j\sigma} c_{j\sigma} c^{\dagger}_{j\bar \sigma}c_{j\bar\sigma}\;,
\end{equation}
which can be approximately factorized as
\begin{equation}
\langle c^\dagger_{1}c_{2}c^\dagger_{3}c_{4}\rangle \approx
\langle c^\dagger_{1}c_{2}\rangle\langle c^\dagger_{3}c_{4}\rangle+\langle c^\dagger_{1}c_{4}\rangle\langle
c_{2}c^\dagger_{3}\rangle\;,
\end{equation}
leading to
\begin{eqnarray}
\langle D_{i}\rangle &\approx& \langle  n_{i\uparrow}\rangle \langle n_{i\downarrow}\rangle = n_{i\uparrow}n_{i\downarrow} \\
\langle N_{i} D_{j}\rangle &\approx& n_{j\sigma}n_{j\bar\sigma}\big(\delta+n_{i\sigma}n_{i\bar\sigma}\big)\nonumber\\
&& +\sum_{\sigma}|\langle c^{\dagger}_{i\sigma} c_{j\sigma}\rangle |^2\big[\tfrac{1}{2}|\langle C^{\dagger}_{i\bar\sigma}
c_{j\bar\sigma}\rangle|^2+n_{j\bar\sigma}(1-n_{i\bar\sigma} )  \big]\;, \label{NIDJ}\nonumber\\
\end{eqnarray}
Noticing that there is a simple relation between $\langle N_{i}\rangle$ and $\langle D_{i}\rangle$:
\begin{eqnarray}
\langle N_{i}\rangle&=&\langle(1- c^{\dagger}_{i\sigma} c_{i\sigma}) (1- c^{\dagger}_{i\bar \sigma}
c_{i\bar\sigma}) \rangle\nonumber\\
&=&1- \sum_{\sigma}\langle c^{\dagger}_{i\sigma} c_{i\sigma} \rangle+\langle  c^{\dagger}_{i\sigma} c
_{i\sigma} c^{\dagger}_{i\bar \sigma} c_{i\bar\sigma}\rangle\nonumber\\
&=&\delta+\langle  D_{i}\rangle\;,
\end{eqnarray}
the holon-doublon  correlator $C_{ij}$ can thus be rewritten as:
\begin{equation}\label{CIJ}
C_{ij} \approx \frac{\sum\limits_{\sigma}|\langle c^{\dagger}_{i\sigma} c_{j\sigma}\rangle |^2\big[\frac{1}{2}|\langle
c^{\dagger}_{i\bar\sigma} c_{j\bar\sigma}\rangle|^2+n_{j\bar\sigma}(1-n_{i\bar\sigma} )\big]}{(\delta+\langle
D_{i}\rangle)\langle D_{j}\rangle}\;.
\end{equation}
We find from Eq.~\eqref{CIJ} that when $U\sim U_{\text{Mott}}$, the decrease of the double occupancy will enhance
$C_{ij}$ dramatically. In the case of $U\gg U_{\text{Mott}}$, the double occupancy decreases at the same rate as
the nearest neighbor hopping probability and more mildly than the next nearest one. Thus, the nearest neighbor
holon-doublon correlator will saturate eventually and the next nearest one decrease after reaching its maximum.

\section{ELECTRON-MOMENTUM DISTRIBUTION}\label{Fermi-Distribution}

Within the cluster slave-spin scheme, the electron momentum distribution $n_d^{(2/4)}(\bm{k}) =\langle
d^\dagger_{\bm{k}\sigma} d_{\bm{k}\sigma}\rangle$ with respect to 2/4-site cluster approximation are as follows.
For the former,
\begin{eqnarray}\label{Fermi-Func-2Site}
n_d^{(2)}(\bm{k})
&\approx& \frac{1}{N} \sum_{j} \langle f_{j\sigma}^\dagger f_{j\sigma} \rangle \nonumber \\
&& + \frac{\langle S_{j+\delta\sigma}^+\rangle\langle S_{j\sigma}^- \rangle}{N} \sum_{j,\delta}
e^{i \bm{k}\cdot \bm{\delta}} \langle f_{j+\delta\sigma}^\dagger
f_{j\sigma} \rangle \nonumber \\
&=&  \frac{1-\delta}{2} -\frac{4Z\epsilon}{t} \gamma_{\bm{k}}\;,
\end{eqnarray}
and for the latter,
\begin{eqnarray}\label{Fermi-Func-4Site}
n_d^{(4)}(\bm{k})&\approx&\frac{1}{N} \sum_{j} \langle f_{j\sigma}^\dagger f_{j\sigma} \rangle \nonumber\\
&& + \frac{\langle S_{j+\delta\sigma}^+ \rangle \langle S_{j\sigma}^- \rangle}{N} \sum_{j,\delta}
e^{i \bm{k}\cdot \bm{\delta}} \langle f_{j+\delta\sigma}^\dagger f_{j\sigma} \rangle\nonumber\\
&& +\frac{\langle S_{j+\eta\sigma}^+\rangle \langle S_{j\sigma}^- \rangle}{N} \sum_{j,\eta}
e^{i \bm{k}\cdot \bm{\eta}} \langle f_{j+\eta\sigma}^\dagger f_{j\sigma} \rangle \nonumber \\
&=& \frac{1-\delta}{2} -\frac{4Z\epsilon}{t} \gamma_{\bm{k}} \nonumber\\
&&+ \frac{4\gamma'_{\bm{k}} }{N}\sum_{\bm{k^\prime}\in \text{RBZ}} \gamma'_{\bm{k^\prime}}
\bigg\{ Z_{\text{ave}} [\theta(-E_{\bm{k^\prime}}^{+}) + \theta(-E_{\bm{k^\prime}}^{-})] \nonumber \\
&& \hspace{-2em} + \frac{Z_\Delta}{\sqrt{(4t Z \gamma_{\bm{k^\prime}})^2 + \Delta^2}}
[\theta(-E_{\bm{k^\prime}}^{+}) - \theta(-E_{\bm{k^\prime}}^{-})] \bigg\}\;,
\end{eqnarray}
where
\begin{eqnarray*}
&& \gamma'_{\bm{k}} = \cos k_x \cos k_y\;, \;\;\;\; E_{\bm{k}}^{\pm}= - \mu^{\text{eff}} \pm \sqrt{(4t
Z \gamma_{\bm{k}})^2 + \Delta_{\sigma}^2} \;, \\
&& Z_{\text{ave}} = \frac{|\langle \tilde{z}_{A\sigma}\rangle|^2 + |\langle \tilde{z}_{B\sigma}\rangle|^2}{2} \;,
\;\;\;\;Z_{\Delta}=\frac{|\langle \tilde{z}_{A\sigma}\rangle|^2 - |\langle \tilde{z}_{B\sigma}\rangle|^2}{2}\;.
\end{eqnarray*}
To understand our present results concerning the negative and beyond unity part of this quantity around $M$
and $\Gamma$ point, we refer to the previous work on slave-particle formalism~\cite{PhysRevB.47.15192},
where a ``no-go'' theorem has been shown to prevent to obtain a complete correct answer for the electron
momentum distribution based merely on a simple decoupling procedure for different degrees of freedom.
In the work, the electron momentum distribution is expressed as:
\begin{equation}\label{Fermi-Feng}
n_c(\bm{k},\sigma)=\frac{1-\delta}{2}+\frac{1}{N}\sum\limits_{\bm{q}}n_{f\sigma}(\bm{k}+\bm{q})n_b(\bm{q})\;.
\end{equation}
The common feature between Eqs.~\eqref{Fermi-Func-2Site}, \eqref{Fermi-Func-4Site}, and \eqref{Fermi-Feng} is
that all of them are corrected from an average occupancy $n_{\sigma}=\frac{1-\delta}{2}$, and it is easy to
check that our results ensure the sum-rule of $N^{-1} \sum_{\bm{k}} n_d(\bm{k}) = \frac{1-\delta}{2}$.

\section{U(1) SLAVE-SPIN FORMALISM IN THE NONINTERACTING LIMIT }\label{Single-Site-Approximation}

In this Appendix, we prove analytically that in the $U(1)$  slave-spin theory~\cite{PhysRevB.86.085104}
and within the single-site approximation, the condition $Z=1$ at $U=0$ recovers the correct noninteracting spinon  dispersion.

Starting from the approximate Hamiltonian of a multiorbital system in the single-site approximation
[Eqs.~(15) and (16) in Yu and Si's work~\cite{PhysRevB.86.085104}
\begin{eqnarray}
H_{f}^{\text{MF}}&=&\sum_{k\alpha\beta}[\epsilon_{k}^{\alpha\beta}\langle\tilde{z}_{\alpha}^{\dagger}\rangle
\langle\tilde{z}_{\beta}\rangle \nonumber \\
&& \;\;\;\;+\delta_{\alpha\beta}(\Delta_{\alpha}+\tilde{\mu}_{\alpha}-\lambda_{\alpha}-\mu)]
f_{k\alpha}^{\dagger}f_{k\beta}\;, \\
H_{S}^{\text{MF}}&=&\sum_{\alpha\beta} \Big[\epsilon^{\alpha\beta}\left(\langle\tilde{z}_{\alpha}^{\dagger}\rangle
\tilde{z}_{\beta}
+\langle\tilde{z}_{\beta}\rangle\tilde{z}_{\alpha}^{\dagger}\right)+\delta_{\alpha\beta}
\frac{\lambda_{\alpha}}{2}(\hat{n}_{\alpha}^{a}-\hat{n}_{\alpha}^{b}) \Big]\nonumber\\
&&+H_{\text{int}}^{S}\;,\label{H-MULTI-ORBIT-S}
\end{eqnarray}
where
\begin{eqnarray}
\tilde{z}_{\alpha}^{\dagger}&=&\langle P_{\alpha}^{+}\rangle a_{\alpha}^{\dagger}b_{\alpha}\langle
P_{\alpha}^{-}\rangle\;,\hspace{2em}\langle P_{\alpha}^{\pm}\rangle=\frac{1}{\sqrt{1/2
\pm(n_{\alpha}^{f}-\frac{1}{2})}}\;,\nonumber\\
\epsilon^{\alpha\beta}&=&\sum_{ij\sigma}t^{\alpha\beta}_{ij}\langle
f^\dagger_{i\alpha\sigma}f_{j\beta\sigma}\rangle\;,\hspace{1.2em}\epsilon^{\alpha\beta}_{\bm{k}}
=\tfrac{1}{N}\sum_{ij}t^{\alpha\beta}_{ij}e^{i\bm{k}\cdot (\bm{r}_{i}-\bm{r}_{j})}\;,\nonumber\\
\tilde{\mu}_{\alpha}&=&2\overline{\epsilon}_{\alpha}\eta_{\alpha}\;,\hspace{7.2em}\eta_{\alpha}
=\frac{2n_{\alpha}^{f}-1}{4n_{\alpha}^{f}(1-n_{\alpha}^{f})}\;,\nonumber\\
\overline{\epsilon}_{\alpha}&=&\sum_{\beta}(\epsilon^{\alpha\beta}\langle\tilde{z}_{\alpha}^{\dagger}\rangle
\langle\tilde{z}_{\beta}\rangle+ \text{c.c.})\;.
\end{eqnarray}
Eq.~\eqref{H-MULTI-ORBIT-S} can be rewritten as
\begin{eqnarray}\label{S-SPIN-MULTI-ORBIT}
H_{S}^{\text{MF}}&=&\sum_{\alpha}[k_{\alpha}\tilde{z}_{\alpha}+k_{\alpha}^{*}\tilde{z}^{\dagger}_{\alpha}
+\frac{\lambda_{\alpha}}{2}(\hat{n}_{\alpha}^{a}-\hat{n}_{\alpha}^{b})]\nonumber\\
&&+H_{\text{int}}^{S}\;,
\end{eqnarray}
where $k_{\alpha}=\sum_{\beta}\epsilon^{\beta\alpha}\langle \tilde{z}^{\dagger}_{\beta}\rangle$. In the
noninteracting limit, because of the decoupling of the orbits, we can keep only one degree of freedom in
Eq.~\eqref{S-SPIN-MULTI-ORBIT}. After dropping the crystal-field splitting $\Delta_{\alpha}$ and the
interaction term, the above Hamiltonians become
\begin{eqnarray}
H_{f}^{\text{MF}}&=&\sum_{k}[\epsilon_{k}\langle\tilde{z}^{\dagger}\rangle\langle\tilde{z}\rangle
+(\lambda+\tilde{\mu}-\mu)]f_{k}^{\dagger}f_{k}\;, \label{MF-HF-1SITE}\\
H_{S}^{\text{MF}}&=&\epsilon\left(\langle\tilde{z}^{\dagger}\rangle\tilde{z}+\langle\tilde{z}\rangle\tilde{z}^{\dagger}
\right)+\frac{\lambda}{2}(\hat{n}^{a}-\hat{n}^{b})\;, \label{MF-H-S-1SITE}
\end{eqnarray}
where $\overline{\epsilon}=2\epsilon\langle\tilde{z}^{\dagger}\rangle\langle\tilde{z}\rangle$ and
$\tilde{\mu}=4\epsilon\eta\langle\tilde{z}^{\dagger}\rangle\langle\tilde{z}\rangle$.

Under the hard-core boson constraint: $a^{\dagger}a+b^{\dagger}b=1$,  the restricted Hillbert space is spanned by
$\{|a=1,b=0\rangle,|a=0,b=1\rangle\}$, in which
the slave-spin Hamiltonian (\ref{MF-H-S-1SITE}) has the form
\begin{eqnarray}
H_{S}^{\text{MF}}&=&\epsilon\left(\langle\tilde{z}^{\dagger}\rangle\tilde{z}+\langle\tilde{z}\rangle\tilde{z}^{\dagger}
\right)+\frac{\lambda}{2}(\hat{n}^{a}-\hat{n}^{b}) \nonumber\\
&=&\left(\begin{array}{c c}\frac{\lambda}{2}&\epsilon R\langle\tilde{z}\rangle\\ \epsilon
R\langle\tilde{z}^{\dagger}\rangle&-\frac{\lambda}{2}\end{array}\right)
\label{MF-H-S-1SITE1}
\end{eqnarray}
with $R=\langle P_{\alpha}^{+}\rangle\langle P_{\alpha}^{-}\rangle$, which has the ground state and energy as
\begin{eqnarray}
|\Psi_{-}\rangle&=&\Big(\tfrac{-\epsilon R\langle\tilde{z}\rangle}{N}\;, \tfrac{\lambda/2+R_{0}}{N} \Big)\;,\\
E_{-}&=&-\sqrt{\tfrac{\lambda^{2}}{2}+\epsilon^{2}R^{2}\lvert\langle\tilde{z}^{\dagger}\rangle\rvert^{2}}\;,
\end{eqnarray}
where $R_{0}=\sqrt{\frac{\lambda^{2}}{4}+\epsilon^{2}R^{2}},\;N=\sqrt{2 R_{0}(\frac{\lambda}{2}+R_{0})}$.
Accordingly, the expectation value of $\hat n_{a}-\hat n_{b} \;\text{and}\; \tilde{z}$ can be calculated as

\begin{equation}
\langle\hat n_{a}-\hat n_{b}\rangle= -\frac{\lambda}{2R_{0}},\;\;\;\; \langle\tilde{z}\rangle= -\frac{\epsilon R^2
\langle\tilde{z}\rangle }{2 R_{0}}\;.\label{SELF-N}
\end{equation}

Because of the constraints $n_{a}-n_{b}=2n^{f}-1$ and $|\langle\tilde{z}\rangle|^{2}=1$ in the noninteracting case, we have
\begin{equation}\label{SELF-Z}
|\langle\tilde{z}\rangle|^{2}=\frac{\epsilon^{2}R^{4}}{4R_{0}^{2}}=1
\end{equation}
with $R=\frac{1}{\sqrt{n^{f}(1-n^{f})}}$. Then, we finally arrive at
\begin{eqnarray}
\lambda&=&-2R_{0}(2n^{f}-1)=\epsilon\frac{2n^{f}-1}{n^{f}(1-n^{f})}\;,\\
\tilde{\mu}&=&4\epsilon\eta=\frac{4\epsilon(2n^{f}-1)}{4n^{f}(1-n^{f})}=\epsilon\frac{(2n^{f}-1)}{n^{f}(1-n^{f})}\;,
\end{eqnarray}
where $\epsilon$ is negative. Thus, $\lambda=\tilde{\mu}$ in the noninteracting case $(Z=1)$  for the single orbital
system, and it is straightforward to generalize the conclusion to the multiorbital systems.

\bibliography{BIBAFM}

\end{document}